\documentclass[12pt,preprint,colordvi]{aastex}
\usepackage{epsf}
\usepackage{color}
\usepackage{amsmath}
\usepackage{amssymb}

\newcommand{\sfrac}[2]{\mathchoice
  {\kern0em\raise.5ex\hbox{\the\scriptfont0 #1}\kern-.15em/
   \kern-.15em\lower.25ex\hbox{\the\scriptfont0 #2}}
  {\kern0em\raise.5ex\hbox{\the\scriptfont0 #1}\kern-.15em/
   \kern-.15em\lower.25ex\hbox{\the\scriptfont0 #2}}
  {\kern0em\raise.5ex\hbox{\the\scriptscriptfont0 #1}\kern-.2em/
   \kern-.15em\lower.25ex\hbox{\the\scriptscriptfont0 #2}}
  {#1\!/#2}}

\def\myhalf {\sfrac{1}{2}}

\def\nph {{n+\myhalf}}

\def\omegadot {{\dot\omega}}
\def\dt       {{\Delta t}}

\def\Fb {{\bf F}}
\def\gb {{\bf g}}
\def\lb {{\bf l}}

\def\Qb {{\bf Q}}

\def\rb {{\bf r}}
\def\Sb {{\bf S}}
\def\ub {{\bf u}}
\def\Ub {{\bf U}}

\def\ellp {{\ell^\prime}}

\def\Hnuc{{H_{\rm nuc}}}

\newcommand{\Code}[1]{\texttt{#1}}

\newcommand{\Cplusplus}{{\rmfamily C\raise.22ex\hbox{\small ++} }}

\def\gcc {{\mathrm{g}~\mathrm{cm}^{-3}}}
\def\cmss {{\mathrm{cm}~\mathrm{s}^{-2}}}
\def\presu {{\mathrm{dyn}~\mathrm{cm}^{-2}}}

\def\Eexpl {\mathcal{E}_{\rm exp}}

\begin{document}

\title{CASTRO: A New Compressible Astrophysical Solver. I. Hydrodynamics and Self-Gravity}

\shorttitle{CASTRO: Hydrodynamics and Self-Gravity}
\shortauthors{Almgren et al.}

\author{A.~S.~Almgren\altaffilmark{1},
        V.~E.~Beckner\altaffilmark{1},
        J.~B.~Bell\altaffilmark{1},
        M.~S.~Day\altaffilmark{1},
        L.~H.~Howell\altaffilmark{2},
        C.~C.~Joggerst\altaffilmark{3},
        M.~J.~Lijewski\altaffilmark{1},
        A.~Nonaka\altaffilmark{1},
        M.~Singer\altaffilmark{2},
        M.~Zingale\altaffilmark{4}}

\altaffiltext{1}{Center for Computational Sciences and Engineering,
                 Lawrence Berkeley National Laboratory,
                 Berkeley, CA 94720}

\altaffiltext{2}{Center for Applied Scientific Computing,
                 Lawrence Livermore National Laboratory,
                 Livermore, CA 94550}

\altaffiltext{3}{Dept. of Astronomy \& Astrophysics,
                 The University of California, Santa Cruz,
                 Santa Cruz, CA 95064;
                 Los Alamos National Laboratory,
                 Los Alamos, CA 87545}

\altaffiltext{4}{Dept. of Physics \& Astronomy,
                 Stony Brook University,
                 Stony Brook, NY 11794-3800}

%==========================================================================
% Abstract
%==========================================================================
\begin{abstract}
We present a new code, CASTRO, that solves the multicomponent compressible hydrodynamic 
equations for astrophysical flows including self-gravity, nuclear reactions 
and radiation. 
CASTRO uses an Eulerian grid and incorporates adaptive mesh refinement (AMR).  
Our approach to AMR uses a nested hierarchy of logically-rectangular grids with 
simultaneous refinement in both space and time.  
The radiation component of CASTRO will be described in detail
in the next paper, Part II, of this series.
\end{abstract}

%------------------------------------------------------------------------
% Introduction
%------------------------------------------------------------------------
\section{Introduction}
In this paper, Part I of a two-part series,
we present a new code, CASTRO, that solves the multicomponent compressible hydrodynamic 
equations with a general equation of state for astrophysical flows.   
Additional physics include self-gravity, nuclear reactions, and radiation.   
CASTRO uses an Eulerian grid and
incorporates adaptive mesh refinement (AMR).  Our approach to AMR uses a 
nested hierarchy of logically-rectangular grids with simultaneous 
refinement of the grids in both space and time.  Spherical (in 1D), 
cylindrical (in 1D or 2D), and Cartesian (in 1D, 2D or 3D) coordinate systems are supported.
The radiation component of CASTRO will be described in detail
in the next paper, Part II, of this series.

There are a number of other adaptive mesh codes for compressible astrophysical flows,
most notably, ENZO \citep{ENZO}, FLASH \citep{flash}, and RAGE \citep{RAGE}.
CASTRO differs from these codes in several ways.   CASTRO uses an
unsplit version of the piecewise parabolic method, PPM, with new limiters that avoid
reducing the accuracy of the scheme at smooth extrema; the other codes
are based on operator-split hydrodynamics, though the  
most recent release of FLASH, version 3.2, includes an unsplit MUSCL-Hancock scheme. 
The different methodologies also vary in their approach to adaptive mesh refinement.
RAGE uses a cell-by-cell refinement strategy while the other codes use patch-based refinement.
FLASH uses equal size patches whereas ENZO and CASTRO allow arbitrary sized patches. 
ENZO and FLASH enforce a strict parent-child relationship between patches; i.e., each refined patch 
is fully contained within a single parent patch; CASTRO requires only that the union of fine patches
be contained within the union of coarser patches with a suitable proper nesting.  
Additionally, FLASH and RAGE use a single time step across all levels while CASTRO and ENZO support
subcycling in time.   All four codes include support for calculation of self-gravity.

It is worth noting that CASTRO uses the same grid structure as the low Mach number astrophysics code, 
%MAESTRO \citep{ABRZ:I, ABRZ:II, ABNZ:III, ZABNW:IV, multilevel}. 
MAESTRO (see, e.g., \cite{multilevel}). 
This will enable us to map the results from a low Mach number simulation, such
as that of the convective period and ignition of a Type Ia supernova, 
to the initial conditions for a compressible simulation such as that of the explosion 
itself, thus taking advantage of the accuracy and efficiency of each approach as appropriate.

\section{Hydrodynamics}
In CASTRO we evolve the fully compressible equations forward in time.  The equations
expressing conservation of mass, momentum, and total energy are:
\begin{eqnarray}
\frac{\partial \rho}{\partial t} &=& - \nabla \cdot (\rho \ub) + S_{{\rm ext},\rho}, \\
\frac{\partial (\rho \ub)}{\partial t} &=& - \nabla \cdot (\rho \ub \ub) - \nabla p + \rho \gb + \Sb_{{\rm ext},\rho\ub}, \\
\frac{\partial (\rho E)}{\partial t} &=& - \nabla \cdot (\rho \ub E + p \ub) 
+ \rho \Hnuc
+ \rho \ub \cdot \gb 
+ S_{{\rm ext},\rho E}.
\end{eqnarray}
Here $\rho$, $\ub$, and $E$ are the mass density, velocity vector, and
total energy per unit mass, respectively.  The total energy, 
$E = e + \ub \cdot \ub / 2$, where $e$ is the specific
internal energy.  The pressure, $p,$ is defined by a
user-supplied equation of state,  and $\gb$ is the gravitational acceleration 
vector.  The source terms, 
$S_{{\rm ext},\rho}$, $S_{{\rm ext},\rho \ub}$,
and $S_{{\rm ext},\rho E}$ are user-specified external source terms for
the mass, momentum, and energy equations, respectively.
For reacting flows, we evolve equations for mass fractions, $X_k$:
\begin{equation}
\frac{\partial (\rho X_k)}{\partial t} = - \nabla \cdot (\rho \ub X_k) + \rho \dot\omega_k + S_{{\rm ext},\rho X_k}.
\end{equation}
where the production rates, 
$\dot\omega_k$, for species $k$ are defined by a user-supplied
reaction network.  The reaction network also determines the energy generation rate $\Hnuc$.
The mass fractions are subject to the constraint that $\sum_k X_k = 1$.
Again, a user-specified
external source, $S_{{\rm ext}, \rho X_k}$, may be specified.
Finally, CASTRO includes passively advected quantities, $C^{\rm adv}_k$, and auxiliary variables,
$C^{\rm aux}_k$ that satisfy 
\begin{eqnarray}
\frac{\partial (\rho C^{\rm adv}_k)}{\partial t} &=& - \nabla \cdot (\rho \ub C^{\rm adv}_k) + S_{{\rm ext},\rho C^{\rm adv}_k} , \\
\frac{\partial (\rho C^{\rm aux}_k)}{\partial t} &=& - \nabla \cdot (\rho \ub C^{\rm aux}_k) + S_{{\rm ext},\rho C^{\rm aux}_k}.
\end{eqnarray}
Advected and auxiliary variables are updated similarly, but they
differ in their usage. In particular, auxiliary variables are passed into the
equation of state routines.  Examples of auxiliary and advected
variables, respectively, might include the electron fraction, $Y_e$,
used in simulations of core collapse supernovae, and angular momentum
in two-dimensional simulations of a rotating star in cylindrical (axisymmetric)
coordinates.  Both of these evolution equations include user-specified
sources, $S_{{\rm ext},\rho C_k^{\rm adv}}$ and $S_{{\rm ext},\rho
  C_k^{\rm aux}}$.  We refer to $\Ub = (\rho,\rho\ub,\rho E,\rho
X_k,\rho C_k^{\rm adv},\rho C_k^{\rm aux})$ as the conserved
variables.

\section{Equation of State and Reaction Network}\label{Sec:EOS Burner}

CASTRO is written in a modular fashion so that the routines for the equation of state 
and reaction network can be supplied by the user.   However, for the
test problems presented later we use routines that come with the CASTRO distribution.

Each equation of state must provide an interface for obtaining thermodynamic quantities
from $\rho, e$, and $X_k$.  One equation of state which is supplied with the 
CASTRO distribution is the gamma-law equation of state, which relates pressure and 
temperature, $T$, to $\rho$ and $e$ via:
\begin{equation}
\label{eq:gamma-law}
p = \rho e (\gamma - 1) = \frac{\rho k_B T}{\mu m_p}.
\end{equation}
Here, $\gamma,$ is the ratio of specific heats
(e.g.\ $\gamma = 5/3$ for a monatomic gas), $k_B$ is Boltzmann's constant, 
$m_p$ is the mass of the proton, and the mean molecular weight, $\mu$, is 
determined by
\begin{equation}
\frac{1}{\mu} = \sum_k  \frac{X_k}{A_k},
\end{equation}
with $A_k$ the atomic weight of species $k$.

The CASTRO distribution also includes more complex equations of state describing
stellar matter, including the Helmholtz equation of state
\citep{timmes_swesty:2000,flash} which includes degenerate/relativistic
electrons, ions (as a perfect gas), and radiation, and the
Lattimer-Swesty equation of state, which describes dense nuclear matter
\citep{LSEOS}.  For tabular
equations of state, it is common that $\rho$, $T$, and $X_k$ are
inputs, in which case a Newton-Raphson iteration is typically used to invert
the equation of state.

CASTRO can support any general reaction network that takes as inputs 
the density, temperature, and mass fractions, and returns updated mass
fractions and the energy release (or decrease).  The input temperature 
is computed from the equation of state before each call to the reaction network.
In general, we expect
the reaction network to evolve the species according to:
\begin{equation}
\frac{dX_k}{dt} = \omegadot_k(\rho,X_k,T).
\end{equation}
Reaction rates can be extremely temperature-sensitive, so in most
cases, the reaction network should be written to evolve the temperature
for the purposes of evaluating the rates.  Close to nuclear
statistical equilibrium, the energy release and change in abundances
can rapidly change sign if the rates are not evaluated 
with a temperature field consistent with the evolving energy \citep{mueller:1986}.
At the end of the burning step, we use the energy release to update 
the total energy, $E$.  The density remains unchanged during the burning.

%It is straightforward to implement additional equation of state and network 
%routines; all that is required is to create an appropriate interface to the 
%CASTRO calls, which can easily be created given the prototypes supplied with the 
%CASTRO distribution.

\section{Gravity}
CASTRO supports several different options for how to specify and/or compute the 
gravitational acceleration.  The simplest option is a gravitational field that 
is constant in space and time; this can be used for small-scale problems in which the 
variation of gravity throughout the computational domain is negligible.
This option is available in 1D Cartesian coordinates, 2D cylindrical or
Cartesian coordinates, and 3D Cartesian coordinates.

A second approach uses a monopole approximation to compute a radial gravity
field consistent with the mass distribution.  Because the algorithm subcycles
in time we construct a separate 1D radial density profile at each level at
each time needed.
%At each level we construct a
%1D radial density profile at the current (or finer) resolution.
%In 1D the radial array aligns with the 
%1D grids so that the average is just a copy;  in 2D and 3D the data from
%individual Cartesian grid cells are added to the correct radial bin and 
%an average is computed over all cells at that radius.  Because in 2D and
%3D the grid boundaries do not align with isocontours of density, data is
%interpolated from coarser levels where needed.
Once the radial density profile is defined, gravity is computed as 
a direct integral of the mass enclosed. This field is then interpolated 
back onto the original grids.  
%In 1D this interpolation is, again, just a copy;
%in 2D and 3D we use quadratic interpolation with limiting so that no new
%maxima or minima are created.

The most general option is to solve the Poisson equation for self-gravity, i.e. solve 
\begin{equation}
\nabla^2\phi = 4\pi G\rho,\label{eq:Self Gravity}
\end{equation}
for $\phi$, and define $\gb = -\nabla\phi.$   
This can be used in any of the coordinate systems.
At boundaries away from the star we set inhomogeneous Dirichlet boundary conditions for $\phi$; these
values are determined by computing the monopole approximation for $\gb$ on the coarsest 
level, integrating this profile radially outward to create $\phi(r),$ and interpolating $\phi$ 
onto the domain boundaries to define the boundary conditions for the solve.  
Boundaries that pass through the center of the star use symmetry boundary conditions.

The Poisson equation is discretized using standard finite difference
approximations and the resulting linear system is solved using geometric
multigrid techniques, specifically V-cycles and red-black Gauss-Seidel relaxation.
For multilevel calculations, special attention is paid to the 
synchronization of the gravitational forcing across levels, which will be
discussed in Section \ref{Sec:AMR}.

There is also an option to add the gravitational forcing due to a 
specified point mass to either of the self-gravity options described above.

\section{Single-Level Integration Algorithm}\label{Sec:SingleLevel}
%We describe the algorithm with gravity computed by solving the Poisson equation;
%modifications for the monopole or constant gravity assumption will be noted.

The time evolution of $\Ub$ can be written in the form
\begin{equation}
\frac{\partial\Ub}{\partial t} = -\nabla\cdot\Fb + \Sb_{\rm react} + \Sb,
\end{equation}
where $\Fb$ is the flux vector, $\Sb_{\rm react}$ are the reaction source terms, 
and $\Sb$ are the non-reaction source terms, 
which includes any user-defined external sources, $\Sb_{\rm ext}$.  
We use Strang splitting \citep{strang:1968} 
to discretize the advection-reaction equations. In other words,
to advance the solution, $\Ub,$ by one time step, $\Delta t,$ we first
advance the nuclear reaction network by $\Delta t / 2$, 
\begin{mathletters}
\begin{equation}
\Ub^{(1)} = \Ub^n + \frac{\dt}{2}\Sb_{\rm react}^n,
\end{equation}
then advect the solution by $\Delta t$, ignoring the reaction terms, 
\begin{equation}
\Ub^{(2)} = \Ub^{(1)} - \Delta t \nabla \cdot\Fb^\nph + \dt\frac{\Sb^{(1)} + \Sb^{(2)}}{2},
\end{equation}
and finally advance the nuclear reaction network by another $\Delta t / 2$, 
\begin{equation}
\Ub^{n+1} = \Ub^{(2)} + \frac{\dt}{2}\Sb_{\rm react}^{(2)} \; .
\end{equation}
\end{mathletters}
The construction of $\Fb$ is purely explicit, and based on an unsplit
Godunov method.  The solution, $\Ub,$ and source terms, $\Sb,$ are defined on cell centers;
we predict the primitive variables, $\Qb = (\rho, \ub, p, \rho e, X_k, C^{\rm adv}_k, C^{\rm aux}_k)$, 
from cell centers at time $t^n$ to edges at time $t^{\nph}$ and use an
approximate Riemann solver to construct fluxes, $\Fb^\nph,$ on cell faces.  
This algorithm is formally second-order in both space and time.

\subsection{Single-Level Flow Chart}\label{Sec:Flow Chart}
At the beginning of each time step, we assume that, 
in the case of self-gravity, ${\gb}$ is defined consistently with the
current mass distribution in $\Ub.$
%(and $\phi$ if appropriate) 
%are defined consistently, i.e., if we are using the Poisson equation to solve
%for gravity, then $\rho^n$ and $\phi^n$ satisfy equation (\ref{eq:Self Gravity}).  
The algorithm at a single level of refinement is composed of the following steps:
\begin{description}
\item[Step 1:] {\em Advance the nuclear reaction network through a time interval of $\Delta t/2$.}

Define $\Ub^{(1)} = \Ub^n$ with the exception of
\begin{eqnarray}
(\rho E)^{(1)} &=& (\rho E)^n + \frac{\dt}{2}(\rho\Hnuc)^n,\\
(\rho X_k)^{(1)} &=& (\rho X_k)^n + \frac{\dt}{2}(\rho\omegadot_k)^n.
\end{eqnarray}
where $(\rho\Hnuc)^n$ and $(\rho\omegadot_k)^n$ are computed using calls to the user-defined
reaction network.  Note that $\rho$ is unchanged during this step.

%\item[Step 2:] {\em Compute gravity at $t^n$.}

%Evaluate gravity using $\rho^{(1)} = \rho^{n}.$   
%\begin{equation}
%\gb^{(1)} = -\nabla\phi^{(1)}, \qquad 
%\Delta\phi^{(1)} = 4\pi G\rho^{(1)},
%\end{equation}

\item[Step 2:] {\em Advect the solution through $\Delta t$.}

Advance the solution using time-centered fluxes and an explicit representation
of the source term, neglecting the contribution from reactions which are taken into
account in {\bf Steps 1} and {\bf 4} (the asterisk superscript notation indicates that
we will later correct this state to effectively time-center the source terms):
\begin{equation}
\Ub^{(2,\ast \ast)} = \Ub^{(1)} - \dt\nabla\cdot\Fb^\nph + \dt\Sb^{(1)}.
\end{equation}

where
\begin{equation}
\Sb_{\Ub}^{(1)} =
\left(\begin{array}{c}
S_{\rho} \\
\Sb_{\rho\ub} \\
S_{\rho E} \\
S_{\rho X_k} \\
S_{\rho C^{\rm adv}_k} \\
S_{\rho C^{\rm aux}_k}
\end{array}\right)^{(1)}
=
\left(\begin{array}{c}
S_{{\rm ext},\rho} \\
(\rho \gb)^{(1)} + \Sb_{{\rm ext},\rho\ub} \\
(\rho \ub \cdot \gb)^{(1)} + S_{{\rm ext},\rho E} \\
S_{{\rm ext},\rho X_k} \\
S_{{\rm ext},\rho C^{\rm adv}_k} \\
S_{{\rm ext},\rho C^{\rm aux}_k}
\end{array}\right)^{(1)}.
\end{equation}
%Note that since the source term is not time-centered, this is not a second-order method.  
%After the advective update, we correct the solution, effectively time-centering the source term. 
The construction of the fluxes is described in detail in Section \ref{Sec:Advection Step}.  We 
note that in the single-level algorithm we can use the gravitational forcing computed in 
{\bf Step 3} of the previous time step, since the density has not changed.

After the advective update, we ensure that the solution is 
physically meaningful by forcing the density to exceed a
non-negative, user-defined minimum value.  We also ensure that the
mass fractions are all non-negative and sum to one.

We also have an option for a user-defined sponge in order to prevent
the velocities in the upper atmosphere from becoming too large, and
subsequently, the time step from becoming too small.  We 
%damp the velocity
%by resetting $\ub^{(2,\ast \ast)}$ using
%\begin{equation}
%\ub^{(2,\ast \ast)} = \frac{\ub^{(2,\ast \ast)}}{1 + \Delta t \kappa f_{\rm damp}(\rho)},
%\end{equation}
multiply the velocity by $1 / (1 + \Delta t \; \kappa \; f_{\rm damp}(\rho)),$
where $\kappa$ is the sponge strength, and $f_{\rm damp}$ is a smooth function of
density that varies from 0 to 1.  
%We say that the sponge is off if $f_{\rm damp} = 0$,
%and that the sponge is at full strength if $f_{\rm damp} = 1$.  
Full details of the sponge are given in \citet{ZABNW:IV}.  
Finally, we adjust $(\rho E)^{(2,\ast \ast)}$ to be consistent with $\ub^{(2,\ast \ast)}$.

\item[Step 3:] {\em Correct the solution with time-centered source terms and 
compute gravity at $t^{n+1}$.}

We correct the solution by effectively time-centering the source terms.  
First, we correct $\Ub$ with updated external sources:  
\begin{equation}
\label{eq:source_update_2}
\Ub^{(2),\ast} = \Ub^{(2,\ast \ast)} + \frac{\dt}{2}\left(S_{{\rm ext},\Ub}^{(2,\ast \ast)} - S_{{\rm ext},\Ub}^{(1)}\right).
\end{equation}

Next, we evaluate gravity using $\rho^{(2, \ast)}.$   If using full gravity we solve
\begin{equation}
\gb^{(2,\ast)} = -\nabla\phi^{(2, \ast)}, \qquad 
\nabla^2 \phi^{(2, \ast)} = 4\pi G\rho^{(2, \ast)},
\end{equation}
where we supply an initial guess for $ \phi^{(2, \ast)}$ from the
previous solve.   In the single-level algorithm described here, 
$\gb^{(2,\ast)}$ is saved to be used as
$\gb^{(1)}$ in {\bf Step 2} of the next time step.
This suffices in the single-level algorithm because $\rho$ does not change between 
the end of {\bf Step 3} of one time step and the start of {\bf Step 2} of the next time step.

We then correct the solution with the updated gravity:
\begin{eqnarray}
(\rho\ub)^{(2)} &=& (\rho\ub)^{(2,\ast)} + \frac{\dt}{2}\left[(\rho\gb)^{(2,\ast)} - (\rho\gb)^{(1)}\right], \\
(\rho E)^{(2)} &=& (\rho E)^{(2,\ast)} + \frac{\dt}{2}
\left[\left(\rho\ub\cdot\gb\right)^{(2,\ast)} - \left(\rho\ub\cdot\gb\right)^{(1)}\right].
\end{eqnarray}

For all other conserved variables other than $\rho\ub$ and 
$\rho E, \Ub^{(2)} = \Ub^{(2,\ast)}$.
We note here that the time discretization of the gravitational forcing terms differs
from that in the FLASH \citep{flash} and ENZO \citep{ENZO} codes, 
where the gravitational forcing at $t^\nph$
is computed by extrapolation from values at $t^n$ and $t^{n-1}$ 
(see also \citealt{bryan:1995}).   Our discretization of the
gravitational terms is consistent with our 
predictor-corrector approach in the handling of other source terms.

\item[Step 4:] {\em Advance the nuclear reaction network through a time interval of $\Delta t/2$.}

%Advance the nuclear reaction network through a time interval of $\Delta t / 2$, i.e., 
Define $\Ub^{n+1} = \Ub^{(2)}$ with the exception of
\begin{eqnarray}
(\rho E)^{n+1} &=& (\rho E)^{(2)} + \frac{\dt}{2}(\rho\Hnuc)^{(2)},\\
(\rho X_k)^{n+1} &=& (\rho X_k)^{(2)} + \frac{\dt}{2}(\rho\omegadot_k)^{(2)}.
\end{eqnarray} 

We also include an option to modify any component of the new-time state as 
needed to account for special user requirements.

\item[Step 5:] {\em Compute the new time step.}

The time step is computed using the standard CFL condition for explicit methods,
with additional constraints (such as one based on rate of burning) possible as needed.  
The user sets a CFL factor, $\sigma^\mathrm{CFL},$
between 0 and 1.  The sound speed, $c$, is computed by the equation of state,
and for a calculation in $n_\mathrm{dim}$ dimensions, 
\begin{equation}
\dt = \sigma^\mathrm{CFL}  \min_{i=1\ldots n_\mathrm{dim}} \left \{ \dt_i \right \},
\end{equation}
where
\begin{equation}
\dt_i = \min_{\bf x}  \left \{ \frac{\Delta x_i}{|{\bf{u}}_i| + c \; }  \right \}.
\end{equation}
$\min_{\bf x}$ is the minimum taken over all computational grid cells in the domain.

\end{description}

This concludes the single-level algorithm description.  We note that whenever
the kinetic energy dominates the total energy, making the calculation of $e$ from 
$E$ numerically unreliable, we use a method similar to the dual-energy approach described 
in \citet{bryan:1995} to compute the internal energy with sufficient precision.  In practice,
this involves evolving $\rho e$ in time and using this solution when appropriate.

%\subsubsection{Dual-Energy Approach}

%In {\bf Steps 1}, {\bf 2}, and {\bf 4}, and after each source term update in
%{\bf Step 3}, we compute an updated value of $\rho E$.  After each of these
%updates, we enforce that $\rho E$ be well-behaved using
%a method similar to the dual-energy approach described in \citet{bryan:1995}.
%For example, at the end of {\bf Step 3},
%\begin{itemize}
%\item If 
%$(\rho e)^{(2)} = (\rho E)^{(2)}-\rho^{(2)}(\ub^{(2)}\cdot\ub^{(2)})/2 < 10^{-4}(\rho E)^{(2)}$,
%we compute a new-time value of $(\rho e)$, which we call $(\rho e)^{\rm corr},$
%using a discretization of the equation
%\begin{equation}
%\frac{\partial (\rho e)}{\partial t} = -\nabla\cdot(\rho\ub e) - p\nabla\cdot\ub 
%+ \rho\Hnuc + S_{{\rm ext},\rho E}.\label{eq:rho e evolution}
%\end{equation}
%\begin{itemize}
%\item If $(\rho e)^{\rm corr} > 0 $ then we redefine 
%$(\rho E)^{(2)} =  (\rho e)^{\rm corr} + \rho^{(2)} (\ub^{(2)} \cdot \ub^{(2)}) / 2$. 
%\item If $(\rho e)^{\rm corr} \leq 0 $ then we compute $e^{\rm corr,2}$ from the equation of
%state using $\rho^{(2)}, X_k^{(2)},$ and a user-defined minimum 
%temperature, and define $(\rho E)^{(2)} =  (\rho e)^{\rm corr, 2} + \rho^{(2)} (\ub^{(2)} \cdot \ub^{(2)}) / 2 $.
%\end{itemize}
%\end{itemize}

\subsection{Construction of Fluxes }\label{Sec:Advection Step}

We use an unsplit Godunov method with characteristic tracing and full
corner coupling in 3D \citep{ppmunsplit} to compute time-centered edge
states.  We have replaced the PPM limiters in \cite{ppmunsplit} with
an updated PPM algorithm that is designed to preserve accuracy at
smooth extrema and is insensitive to asymmetries caused by roundoff
error \citep{ppm2,ppm3}.  CASTRO also has options to use the unsplit
piecewise-linear algorithm described in \cite{colella1990,saltzman}, 
or to retain the PPM limiters in \cite{ppmunsplit}, which were 
originally developed in \cite{ppm} using a split integrator.

There are four major steps in the construction of the face-centered
fluxes, $\Fb^{\nph}$ that are used in {\bf Step 2} in Section
\ref{Sec:Flow Chart} to update the solution.  We also include details
on the solution of the Riemann problem.  In summary,

\quad{\bf Step 2.1:} Rewrite the state, $\Ub^{(1)}$, in terms of primitive variables, $\Qb^{(1)}$.

\quad{\bf Step 2.2:} Construct a piecewise parabolic approximation of $\Qb^{(1)}$ within each cell.

\quad{\bf Step 2.3:} Predict average values of $\Qb^{(1)}$ on edges over the time step using 
characteristic extrapolation.

\quad{\bf Step 2.4:} Compute fluxes,  $\Fb^{\nph},$ using an approximate Riemann problem solver.

We expand each of these steps in more detail below.

\begin{description}
\item[Step 2.1:] {\em Compute primitive variables and source terms.}

We define $\Qb^{(1)} = (\rho, \ub, p, \rho e, X_k, C^{\rm adv}_k,
C^{\rm aux}_k)^{(1)}$.  The pressure, $p$, is computed through a call
to the equation of state using $\rho, e$, and $X_k$. Note that we also
include $\rho e$ in $\Qb$; this quantity is used in the approximate
Riemann solver to avoid an EOS call to evaluate the energy flux,
analogous to the effective dynamics for $\gamma = p/(\rho e) + 1$ 
in the \cite{colellaglaz1985} approximate Riemann solver.

For the overall integration algorithm, we want to include the effect
of source terms except for reactions in the characteristic tracing
(Step 2.2). (Reactions are treated separately using a symmetric
operator split approach in {\bf Steps 1} and {\bf 4} of the algorithm.)  The time
evolution equations written in terms of the primitive variables,
$\Qb,$ and omitting contributions from reactions, are

\begin{equation}
\frac{\partial \Qb}{\partial t} =
\left(\begin{array}{c}
 -\ub\cdot\nabla\rho - \rho\nabla\cdot\ub \\
 -\ub\cdot\nabla\ub - \frac{1}{\rho}\nabla p \\
 -\ub\cdot\nabla p - \rho c^2\nabla\cdot\ub \nonumber \\
 - \ub\cdot\nabla(\rho e) - (\rho e+p)\nabla\cdot\ub \\
 -\ub\cdot\nabla X_k \\
 -\ub\cdot\nabla C^{\rm adv}_k \\
 -\ub\cdot\nabla C^{\rm aux}_k 
\end{array}\right)
+ \Sb_{\Qb}
\end{equation}
where  
\begin{equation}
\Sb_{\Qb} =
\left(\begin{array}{c}
S_\rho \\
\Sb_{\ub} \\
S_p \\
S_{\rho e} \\
S_{X_k} \\
S_{C^{\rm adv}_k} \\
S_{C^{\rm aux}_k}
\end{array}\right)
=
\left(\begin{array}{c}
S_{{\rm ext},\rho} \\
\gb + \frac{1}{\rho}\Sb_{{\rm ext},\rho\ub} \\
%\frac{p_e}{\rho}\left(\rho\Hnuc + S_{{\rm ext},\rho E}\right) + p_\rho S_{{\rm ext}\rho} + \frac{p_{X_k}}{\rho}\left(\rho\omegadot_k + S_{{\rm ext},\rho X_k}\right)\\
\frac{p_e}{\rho} S_{{\rm ext},\rho E} + p_\rho S_{{\rm ext}\rho} + \frac{p_{X_k}}{\rho} S_{{\rm ext},\rho X_k}\\
%\rho\Hnuc + S_{{\rm ext},\rho E} \\
\ S_{{\rm ext},\rho E} \\
\frac{1}{\rho}S_{{\rm ext},\rho X_k} \\
\frac{1}{\rho}S_{{\rm ext},\rho C^{\rm adv}_k} \\
\frac{1}{\rho}S_{{\rm ext},\rho C^{\rm aux}_k}
\end{array}\right)
\end{equation}
Here, $c$ is the sound speed,
defined as $c = \sqrt{\Gamma_1 p /\rho}$, with $\Gamma_1 = d\log
p/d\log \rho |_s$, with $s$ the entropy.  The remaining thermodynamic
derivatives are $p_e = \partial p / \partial e |_{\rho, X_k}$, $p_\rho
= \partial p / \partial \rho |_{e, X_k}$, and $p_{X_k} = \partial p /
\partial X_k |_{\rho, e, X_{j,(j\ne k)}}$.  Often, the equation
of state is a function of $\rho$, $T$, and $X_k$, and returns derivatives
with these quantities held constant.  In terms of the latter derivatives, 
our required thermodynamic derivatives are:
\begin{mathletters}
\begin{eqnarray*}
p_e &=&
\left ( \left . \frac{\partial e}{\partial T} \right |_{\rho, X_k} \right )^{-1}
        \left . \frac{\partial p}{\partial T} \right |_{\rho,X_k}, \\
p_\rho &=& 
\left . \frac{\partial p}{\partial \rho} \right |_{T,X_k} -
\left ( \left . \frac{\partial e}{\partial T} \right |_{\rho, X_k} \right )^{-1}
        \left . \frac{\partial p}{\partial T} \right |_{\rho,X_k}
        \left . \frac{\partial e}{\partial \rho} \right |_{T,X_k}, \\
p_{X_k} &=& 
\left . \frac{\partial p}{\partial X_k} \right |_{\rho,T,X_{j,(j\ne k)}} -
\left ( \left . \frac{\partial e}{\partial T} \right |_{\rho, X_k} \right )^{-1}
        \left . \frac{\partial p}{\partial T} \right |_{\rho,X_k}
        \left . \frac{\partial e}{\partial X_k} \right |_{\rho,T,X_{j,(j \ne k)}}.
\end{eqnarray*}
\end{mathletters}

%Note that we neglect reaction source terms since they are accounted for in {\bf Steps 1} and {\bf 5}.  

%\begin{itemize}
%\item $\rho$ - directly copy from the conserved state vector.
%\item $\rho e$ - compute using $\rho e = \rho(E-\ub\cdot\ub/2)$.  Note that 
%we have a similar option to protect against negative values of $e$ as described in 
%Section \ref{Sec:EOS Burner}.
%\item $\ub, T_k, X_k, Y_k$ - copy these from the conserved state vector, dividing by $\rho$.
%\item $p,T$ - use the equation of state to compute $p,T = p,T(\rho,e,X_k)$.  
%\end{itemize}

\item[Step 2.2:] {\em Reconstruct parabolic profiles within each cell.}

%\begin{eqnarray}
%\frac{\partial\rho}{\partial t} &=& -\ub\cdot\nabla\rho - \rho\nabla\cdot\ub + S_{{\rm ext},\rho}, \\
%\frac{\partial\ub}{\partial t} &=& -\ub\cdot\nabla\ub - \frac{1}{\rho}\nabla p + \gb + \frac{1}{\rho}\Sb_{{\rm ext},\rho\ub}, \\
%\frac{\partial p}{\partial t} &=& -\ub\cdot\nabla p - \rho c^2\nabla\cdot\ub \nonumber \\
%&&+ \frac{p_e}{\rho}\left(\rho\Hnuc + S_{{\rm ext},\rho E}\right) + p_\rho S_{{\rm ext},\rho} + \frac{p_{X_k}}{\rho}\left(\rho\omegadot_k + S_{{\rm ext},\rho X_k}\right), \\
%\frac{\partial(\rho e)}{\partial t} &=& - \ub\cdot\nabla(\rho e) - (\rho e+p)\nabla\cdot\ub + \rho\Hnuc + S_{{\rm ext},\rho E}, \\
%\frac{\partial X_k}{\partial t} &=& -\ub\cdot\nabla X_k + \dot\omega_k + \frac{1}{\rho}S_{{\rm ext},\rho X_k}, \\
%\frac{\partial C^{\rm adv}_k}{\partial t} &=& -\ub\cdot\nabla C^{\rm adv}_k + \frac{1}{\rho}S_{{\rm ext},\rho C^{\rm adv}_k}, \\
%\frac{\partial C^{\rm aux}_k}{\partial t} &=& -\ub\cdot\nabla C^{\rm aux}_k + \frac{1}{\rho}S_{{\rm ext},\rho C^{\rm aux}_k}.
%\end{eqnarray}
%where $p_e=\partial p/\partial e|_{\rho,X_k}$, $p_\rho=\partial p/\partial\rho|_{e,X_k}$, $p_{X_k}=\partial p/\partial X_k|_{\rho,e,(X_j,j\ne k)}$, and $c$ is the sound speed, 
%\begin{equation}
%c = \sqrt{\frac{\Gamma_1 p}{\rho}},
%\end{equation}
%where $\Gamma_1 = d \log p / d\log \rho$ at constant entropy.  Note that for the gamma-law equation of state, $\Gamma_1 = \gamma$.

In this step we construct a limited piecewise parabolic profile of each $q$ in $\Qb$ 
(we use $q$ to denote an arbitrary primitive variable from from $\Qb$).
These constructions are performed in each coordinate direction separately.  
The default option in CASTRO is to use a new limiting procedure that avoids reducing the 
order of the reconstruction at smooth local extrema.
The details of this construction are given in \cite{ppm2,ppm3}.  In summary:
\begin{itemize}
%\item {\bf Step 2.2a}: For each cell, compute $q_{i,+}$ and $q_{i,-}$, which are
%spatial interpolations, with limiting, of $q^n$ to the high and low faces of cell $q_i$,
%respectively.
\item {\bf Step 2.2a}: For each cell, we compute the spatial interpolation of $q^n$ to the
high and low faces of cell $q_i$ using a limited cubic interpolation formula.
These interpolants are denoted by $q_{i,+}$ and $q_{i,-}$.
\item {\bf Step 2.2b}: Construct quadratic profiles using $q_{i,-},q_i$, and $q_{i,+}$.
\begin{equation}
q_i^{\rm quad}(x) = q_{i,-} + \xi(x)\left\{q_{i,+} - q_{i,-} + q_{6,i}[1-\xi(x)]\right\},\label{Quadratic Interp}
\end{equation}
\begin{equation}
q_6 = 6q_{i} - 3\left(q_{i,-}+q_{i,+}\right),
\end{equation}
\begin{equation}
\xi(x) = \frac{x - ih}{h}, ~ 0 \le \xi(x) \le 1 \; ,
\end{equation}
where $h$ is the mesh spacing in the direction of interpolation.
Also, as in \cite{ppmunsplit}, we compute a flattening coefficient, $\chi\in[0,1]$, 
used in the edge state prediction to further limit slopes near strong shocks.  
The computation of $\chi$ is identical to the approach used in  FLASH \citep{flash}, 
except that a flattening 
coefficient of 1 indicates that no additional limiting takes place, whereas a 
flattening coefficient of 0 means we effectively drop order to a first-order Godunov 
scheme, which is opposite of the convention used in FLASH.
\end{itemize}

\item[Step 2.3:] {\em Characteristic extrapolation.}

We begin by extrapolating $\Qb^{(1)}$ to edges at $t^{\nph}.$
The edge states are dual-valued, i.e., at each face, there is a left state and a right 
state estimate, denoted $q_{L,i+\myhalf}$ and $q_{R,i+\myhalf}$ (we write the equations in 1D 
for simplicity).  The spatial extrapolation is one-dimensional, i.e., transverse 
derivatives are omitted and accounted for later.
\begin{itemize}
\item {\bf Step 2.3a:} Integrate the quadratic profiles.  We are essentially computing the 
average value swept out by the quadratic profile across the face assuming the profile 
is moving at a speed $\lambda_k$, where $\lambda_k$ is a standard wave speed associated with gas dynamics.\\ \\
Define the following integrals, where $\sigma_k = |\lambda_k|\Delta t/h$:
\begin{mathletters}
\begin{eqnarray}
\mathcal{I}_{i,+}(\sigma_k) &=& \frac{1}{\sigma_k h}\int_{(i+\myhalf)h-\sigma_k h}^{(i+\myhalf)h}q_i^{\rm quad}(x)dx \\
\mathcal{I}_{i,-}(\sigma_k) &=& \frac{1}{\sigma_k h}\int_{(i-\myhalf)h}^{(i-\myhalf)h+\sigma_k h}q_i^{\rm quad}(x)dx
\end{eqnarray}
\end{mathletters}
Substituting (\ref{Quadratic Interp}) gives:
\begin{mathletters}
\begin{eqnarray}
\mathcal{I}_{i,+}(\sigma_k) &=& q_{i,+} - \frac{\sigma_k}{2}\left[q_{i,+}-q_{i,-}-\left(1-\frac{2}{3}\sigma_k\right)q_{6,i}\right], \\
\mathcal{I}_{i,-}(\sigma_k) &=& q_{i,-} + \frac{\sigma_k}{2}\left[q_{i,+}-q_{i,-}+\left(1-\frac{2}{3}\sigma_k\right)q_{6,i}\right].
\end{eqnarray}
\end{mathletters}
\item {\bf Step 2.3b:} Obtain a left and right edge state at $t^\nph$
   by applying a characteristic tracing operator (with flattening) to the integrated
  quadratic profiles.  Note that we also include the explicit source
  term contribution.
\begin{mathletters}
\begin{eqnarray}
q_{L,i+\myhalf} &=& q_i - \chi_i\sum_{k:\lambda_k \ge 0}\lb_k\cdot\left[q_i-\mathcal{I}_{i,+}(\sigma_k)\right]\rb_k + \frac{\dt}{2}S_{q,i}^n, \\
q_{R,i-\myhalf} &=& q_i - \chi_i\sum_{k:\lambda_k \le 0}\lb_k\cdot\left[q_i-\mathcal{I}_{i,-}(\sigma_k)\right]\rb_k + \frac{\dt}{2}S_{q,i}^n.
\end{eqnarray}
\end{mathletters}
In non-Cartesian coordinates, volume source terms are added to the traced states.
Here, $\rb_k$ and $\lb_k$ are the standard right column and left row eigenvectors 
associated with the equations of gas dynamics (see \citealt{Toro}).

An unsplit approximation that includes full corner coupling is
constructed by constructing increasingly accurate approximations to the
transverse derivatives.  The details follow exactly as given in
Section 4.2.1 in \cite{ppmunsplit}, except for the solution of the
Riemann problem, which is described in {\bf Step 2.4}.
\end{itemize}

\item[Step 2.4:] {\em Compute fluxes}

The fluxes are computed using an approximate Riemann solver.  
The solver used here is essentially the same as that used in \citet{cgf}, 
which is based on ideas discussed in \citet{bellcolellatrangenstein}.
This solver is computationally faster and considerably simpler 
than the approximate Riemann solver introduced by \citet{colellaglaz1985}.  
The Colella and Glaz solver was based on an effective
dynamics for $\gamma$ and was designed for real gases  
that are well-approximated by this type of model. 
The approximate Riemann solver used in CASTRO is suitable
for a more general convex equation of state.  

As with other approximate Riemann solvers, an important design principle is to avoid additional evaluations
of the equation of state when constructing the numerical flux.
For that reason, we include $\rho e$ in $\Qb$ and compute $(\rho e)_{L,R}$.
The information carried in $\rho e$ is overspecified but it allows us to compute an energy flux without
an inverse call to the equation of state.

The numerical flux computation is based on approximating the solution to the Riemann problem and evaluating
the flux along the $x/t = 0$ ray. The procedure is basically a two-step process in which we first
approximate the solution in phase space and then interpret the phase space solution in real space.

\begin{itemize}
\item {\bf Step 2.4a}: 
To compute the phase space solution we first solve for $p^*$ and $u^*$, the pressure between
the two acoustic waves and the velocity of the contact discontinuity, respectively.  
These quantities are computed using a linearized approximation to the Rankine-Hugoniot relations.
We first define $\Gamma_{1,L/R}$ by using the cell-centered values on either side of the
interface.
Next, we compute Lagrangian sound speeds, $W_L = \sqrt { \Gamma_{1,L} p_L \rho_L}$ and
$W_R = \sqrt { \Gamma_{1,R} p_R \rho_R}$ and 
the corresponding Eulerian sound speeds 
$c_{L,R} = \sqrt { \Gamma_{1,L,R} p_{L,R} / \rho_{L,R}}$.  Then,
\begin{mathletters}
\begin{equation}
p^* = \frac{W_L p_R + W_R p_L + W_L W_R(u_L-u_R)}{W_L + W_R},
\end{equation}
\begin{equation}
u^* = \frac{W_L u_L + W_R u_R+ (p_L - p_R)}{W_L + W_R}.
\end{equation}
\end{mathletters}
%See, for example, \citealt{Toro}, chpt 9 for a derivation of these expressions.
%\MarginPar{I added something but one shouldn't give Toro credit for this. These expressions
%are from Colella et al. 1997 reference above based on Bell et al. cite above as well.  Does this
%seem OK?}
From $u^*$ and $p^*$ we can compute 
\begin{mathletters}
\begin{eqnarray}
\rho^*_{L,R} &=& \rho_{L,R} + \frac{p^* - p_{L,R}}{c^2_{L,R}}, \\
(c^*_{L,R})^2 &=& \Gamma_{1,L,R} p^*_{L,R} / \rho^*_{L,R}, \\
(\rho e)^*_{L,R} &=& (\rho e)_{L,R} + (p^* - p_{L,R})\frac{(e+p/ \rho)_{L,R}}{c^2_{L,R}}, \\
v^*_{L,R} &=& v_{L,R},
\end{eqnarray}
\end{mathletters}
where $v$ generically represents advected quantities (which includes transverse velocity components).
Here, the notation $^*_{L,R}$ refers to values on the left and right side of the contact discontinuity.

\item {\bf Step 2.4b}: 
The next step in the approximate Riemann solver is to interpret this
phase space solution.  
If $u^*>0$ then the contact discontinuity is moving to the right and
numerical flux depends on the speed and structure of the acoustic wave
connecting $\Qb_L$ and $\Qb^*_L$ associated with the $\lambda = u-c$
eigenvalue.  Similarly if $u^*<0$ then the contact is moving to the
left and the numerical flux depends on the speed and structure of the
acoustic wave connecting $\Qb_R$ and $\Qb^*_R$ associated with the
$\lambda = u+c$ eigenvalue.  Here we discuss in detail the case in
which $u^*>0;$ the other case is treated analogously.

For $u^*>0$ we define $\lambda_L = u_L - c_L$ and $\lambda^*_L = u^*_L - c^*_L$. 
If $p^*_L > p_L$ then the wave is a shock wave and we define a shock speed 
$\sigma = \myhalf (\lambda_L + \lambda^*_L)$.
For that case if $\sigma > 0$ then the shock is moving to the right and we define 
the Godunov state $\Qb_G = \Qb_L$; otherwise $\Qb_G = \Qb^*_L$.
The rarefaction case is somewhat more complex. 
If both $\lambda_L$ and $\lambda^*_L$ are negative, then the rarefaction fan is 
moving to the left and $\Qb_G = \Qb^*_L$.
Similarly, if both  $\lambda_L$ and $\lambda^*_L$ are positive, then the 
rarefaction fan is moving to the right and $\Qb_G = \Qb_L$.
However, in the case in which $\lambda_L < 0 <  \lambda^*_L$, the rarefaction 
spans the $x/t=0$ ray and we need to interpolate the solution. For this case, we define
\begin{equation}
\Qb_G = \alpha \Qb^*_L + (1-\alpha) \Qb_L
\end{equation}
where $\alpha = \lambda_L / (\lambda_L - \lambda^*_L)$. 
This choice of $\alpha$ corresponds to linearly interpolating $\Qb$ through the 
rarefaction to approximate the state that propagates with zero speed.

As noted above the case in which $u^*<0$ is treated analogously. When $u^* =0$ we 
compute $\Qb_G$ by averaging $\Qb^*_L$ and $\Qb^*_R$.
For the Riemann problem approximation, we allow for user-specified floors for 
$\rho$, $p$ and $c$ to prevent the creation of non-physical values. 
\end{itemize}

The fluxes can then be evaluated from the final $\Qb_G$.  
A small quadratic artificial viscosity that is proportional to the 
divergence of the velocity field is added to the flux in order to add
additional dissipation at strong compressions.  We also scale all the
species fluxes so that they sum to the density flux, as in the sCMA
algorithm described by \cite{PlewaMueller:1999}. 

%Because the limiters are applied on a component-by-component basis it is possible
%that the species no longer sum to one on faces.   Before constructing fluxes for the species
%advection we apply a correction described as the 
%We define a correction factor as the inverse of the sum of the species on the face; we then
%multiply the value of each species on the edge by this factor; the resultant sum is 
%now identically one.

\end{description}

\section{AMR}\label{Sec:AMR}
Our approach to adaptive mesh refinement in CASTRO uses a nested
hierarchy of logically-rectangular grids with simultaneous refinement
of the grids in both space and time.  The integration algorithm on the grid hierarchy
is a recursive procedure in which coarse grids are advanced in time,
fine grids are advanced multiple steps to reach the same time
as the coarse grids and the data at different levels are then synchronized.

The AMR methodology was introduced by \cite{berger-oliger};
it has been demonstrated to be highly successful for gas dynamics by 
\cite{berger-colella} in two dimensions and by \cite{bell-3d} in three dimensions.

\subsection{Creating and Managing the Grid Hierarchy}

\subsubsection{Overview}

\begin{figure}[t]
\centering
\setlength{\unitlength}{0.25cm}
\includegraphics{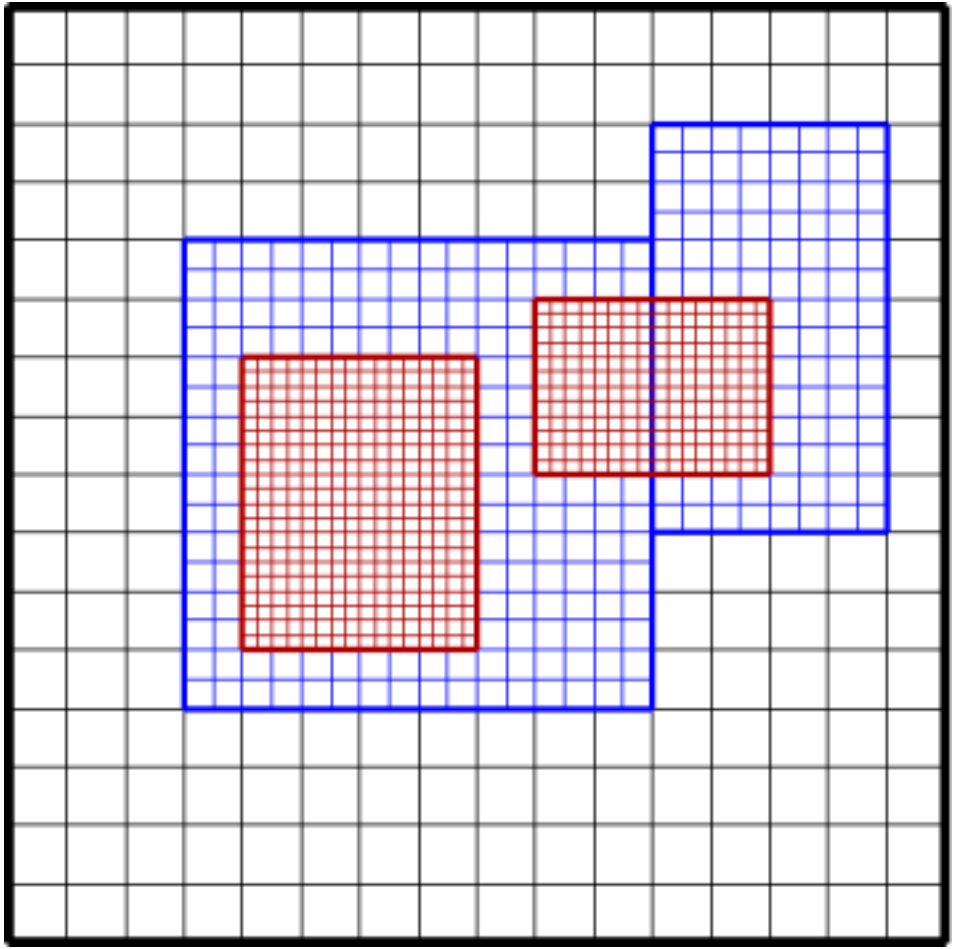}
\caption{\label{fig:grid_cartoon} Cartoon of AMR grids with two levels of factor 2 refinement.  
The black grid covers the domain with $16^2$ cells.  Bold lines represent grid boundaries,
the different colors represent different levels of refinement. 
The two blue grids are at level 1 and the cells
are a factor of two finer than those at level 0.  The two red grids are at level 2 and the
cells are a factor of two finer than the level 1 cells.   Note that the level 2 grids
are properly nested within the union of level 1 grids, but there is no direct parent-child connection.}
\end{figure}

The grid hierarchy is composed of different levels of refinement ranging
from coarsest ($\ell = 0$) to finest ($\ell = {\ell}_{\rm finest}$).
The maximum number of levels of refinement allowed, $\ell_{\rm max}$, is specified at the start 
of a calculation. At any given time in the calculation 
there may not be that many levels in the hierarchy, i.e.\ $\ell_{\rm finest}$
can change dynamically as the calculation proceeds as long as 
$\ell_{\rm finest} \leq \ell_{\rm max}.$ 
Each level is represented by the union of non-overlapping rectangular grids 
of a given resolution.  Each grid is composed of an even
number of cells in each coordinate direction; cells 
are the same size in each coordinate direction but grids
may have different numbers of cells in each direction.  Figure~\ref{fig:grid_cartoon}
shows a cartoon of AMR grids in two dimensions with two levels of refinement.

In this implementation, the refinement ratio between levels
$\ell$ and $\ell+1,$ which we call $r_\ell,$ is always two or four,
with the same factor of refinement in each coordinate direction.
%i.e.\ $\Delta x^{\ell} = \Delta y^{\ell}
%= \Delta z^{\ell} = r_{\ell}^{\ell+1} \Delta x^{\ell+1}$
%for $\ell < \ell_{\rm finest}$,
%where $r$ is the refinement ratio between levels $\ell$ and $\ell+1$.
%(We note here that neither isotropic refinement nor uniform base
%grids are requirements of the fundamental algorithm) 
%In the actual implementation, the
%refinement ratio, either 2 or 4, can be a function of level;
%however, in the exposition we will assume that $r_{\ell}^{\ell+1}$ is constant
%for all $\ell=[0,\ell_{{\rm finest}-1}]$.
The grids are properly nested, in the sense that the union of grids
at level $\ell+1$ is contained in the union of grids at level $\ell$.
Furthermore, the containment is strict in the sense that,
except at physical boundaries,
the level $\ell$ grids are large enough to guarantee that there is
a border at least $n_{\rm proper}$ level $\ell$ cells wide surrounding each level
$\ell +1$ grid (grids at all levels are allowed to extend to the physical
boundaries so the proper nesting is not strict there).  The parameter
$n_{\rm proper}$ is two for factor two refinement, and one for factor four
refinement, since four ghost cells are needed for the PPM algorithm.

\subsubsection{Error Estimation and Regridding}

We initialize the grid hierarchy and regrid following the procedure 
outlined in \cite{bell-3d}.
%The grid hierarchy is constructed using an error estimation criterion
%to determine where additional resolution is required.  
Given grids at level $\ell$ we use an error estimation
procedure to tag cells where the error, as defined by user-specified
routines, is above a given tolerance.   Typical error criteria include
first or second derivatives of the state variables or quantities
derived from the state variables, or the state variables or 
derived quantities themselves.   A user can specify that any or all of the
criteria must be met to refine the cell;  one can also specify criteria that
ensure that a cell not be refined.  For example, one could specify that
a cell be refined if $\rho > \rho_{\rm crit}$ and
( $ (\nabla^2 T) > (\nabla ^2 T)_{\rm crit}$ or $ |\nabla p| > |\nabla p|_{\rm crit}$ ),
where $\rho_{\rm crit},  (\nabla ^2 T)_{\rm crit},$ and $|\nabla p|_{\rm crit}$
are constants specified by the user.

The tagged cells are grouped into rectangular grids at level $\ell$ using
the clustering algorithm given in \cite{bergerRigoutsos:1991}.
These rectangular patches are refined to form the grids at level $\ell+1$.
Large patches are broken into smaller patches for distribution to multiple 
processors  based on a user-specified
{\it max\_grid\_size} parameter.
%The process is repeated until either an
%error tolerance criterion is satisfied or a specified maximum level is reached.  
%Finally, we impose the proper nesting requirements.

At the beginning of every $k_\ell$ level $\ell$ time steps,
where $k_\ell \geq 1$ is specified by the user at run-time,
new grid patches are defined at all levels $\ell+1$ and higher
if $\ell < \ell_{\rm max}.$   In regions previously covered by 
fine grids the data is simply copied from old grids to new; in 
regions which are newly refined, data is interpolated from underlying
coarser grids.

\subsubsection{Enlarging the Domain}

The finest resolution of a calculation can vary in time;
however, the coarsest resolution covering the domain does not 
change during a single run.  However, 
a feature has been added to the CASTRO distribution 
that allows a user to restart a calculation in a larger domain covered by a 
coarser resolution, provided the data exists to initialize the larger domain.
This is useful in simulations during which a star expands dramatically, 
for example.   Using this strategy one could periodically stop the simulation,
double the domain size, and restart the calculation in the larger domain.

\subsection{Multilevel Algorithm}

\subsubsection{Overview}

The multilevel time stepping algorithm can most easily be thought of
as a recursive procedure.  In the case of zero or constant gravity, 
to advance level $\ell,$ $0 \leq \ell \leq \ell_{\rm max}$ the following steps are taken.
Here the phrase, ``Advance $\Ub$'' refers to {\bf Steps 1--4} of the single-level
algorithm described in the previous section.   

\begin{itemize}
\item 
If $\ell=0,$ compute the new time steps for all levels as follows
\begin{itemize}
\item compute the appropriate time step for each level, $\Delta t^{\ellp,*}$
using the procedure described in {\bf Step 5} of the previous section,
\item define $R_\ellp$ as the ratio of the level 0 cell size to the level $\ellp$ cell size
\item define $\Delta t^{0} = {\bf min}_\ellp ( R_\ellp \Delta t^{\ellp,*}),$
\item define $\Delta t^{\ellp} = \Delta t^0 / R_\ellp$ for all $\ellp$,  $0 \leq \ellp \leq \ell_{\rm max}$
\end{itemize}

\item{Advance $\Ub$ at level $\ell$ in time as if it is the only level,
filling boundary conditions for $\Ub$ from level $\ell-1$ if level $\ell > 0$,
and from the physical domain boundaries.}

\item{If $\ell < \ell_{max}$
\begin{itemize}
\item{Advance $\Ub$ at level ($\ell+1$) for $r_\ell$ time steps 
with time step $\Delta t^{\ell+1} = \frac{1}{r_\ell} \Delta t^{\ell}.$}

\item{Synchronize the data between levels $\ell$ and $\ell+1$}

\begin{itemize}
\item Volume average $\Ub$ at level $\ell+1$ onto level $\ell$ grids.
\item Correct $\Ub$ in all level $\ell$ cells adjacent to but not covered by the union of
level $\ell+1$ grids through an explicit refluxing operation as described in
\cite{berger-colella}.
\end{itemize}
\end{itemize}
}
\end{itemize}

\subsubsection{Monopole Gravity}

When we use the monopole gravity assumption in a multilevel simulation, we can no longer exploit the
fact that $\rho$ at level $\ell$ at the end of {\bf Step 3} of one time step is unchanged
when one reaches the beginning of {\bf Step 2} of the next level $\ell$ time step.  If $\ell <  \ell_{max},$
then potential changes in $\rho$ come from two sources:
\begin{itemize}
\item $\rho$ at level $\ell$ under the level $\ell+1$ grids is replaced by 
the volume average of $\rho$ at level $\ell+1$;
\item the explicit refluxing step between levels $\ell$ and $\ell+1$ modifies $\rho$ 
on all level $\ell$ cells adjacent to but not covered by the union of
level $\ell+1$ grids.
\end{itemize}

In addition, because the grids are dynamically created and destroyed through regridding,
at the beginning of {\bf Step 2} of a level $\ell$ time step, there may not be a
value for $\gb$ from the previous step, because this region of space was 
previously not covered by level $\ell$ grids. 

In order to address all of these changes, we simply compute $\gb^{(1)}$ at 
the beginning of {\bf Step 2} of each time step at each level, rather than copying
it from $\gb^{(2,*)}$ from {\bf Step 3} of the previous time step as in the single-level algorithm.
This captures any changes in grid structure due to regridding, and reflects any changes in density
due to refluxing or volume averaging.  

\subsubsection{Full Gravity Solve}

\subsubsubsection{Overview}

Solving the Poisson equation for self-gravity on a multilevel grid
hierarchy introduces additional complications.   We start by defining
some necessary notation.
%We define here two types of solves used
%for the equation 
%\[ \nabla^2 \phi =  4\pi G\rho . \]
We define $L^\ell$ as an approximation to $\nabla^2$ at level $\ell,$ with the
assumption that Dirichlet boundary conditions are supplied on the boundary of the
union of level $\ell$ grids (we allow more general boundary conditions at physical
boundaries), and define a {\it level solve} as the process of solving 
\[ L^{\ell} \phi^{\ell} =  4\pi G\rho^{\ell} \]
at level $\ell.$

We define $L_{\ell,m}^{\rm comp}$ as the composite grid approximation
to $\nabla^2$ on levels $\ell$ through $m$, and define a {\it composite solve}
as the process of solving 
\[ L_{\ell,m}^{\rm comp} \phi^{\rm comp} =  4\pi G\rho^{\rm comp} \]
on levels $\ell$ through $m.$
The solution to the composite solve satisfies 
\[ L^{m} \phi^{\rm comp} =  4\pi G\rho^{m} \]
at level $m$, but satisfies
\[ L^{\ellp} \phi^{\ellp} =  4\pi G\rho^{\ellp} \]
for $\ell \leq \ellp < m$ only on the regions of each level
{\it not} covered by finer grids or adjacent to the boundary 
of the finer grid region.  In regions of a level $\ellp$ grid
covered by level $\ellp+1$ grids
the solution is defined as the volume average of the solution at $\ellp+1$;
in level $\ellp$ cells immediately adjacent to the boundary
of the union of level $\ellp+1$ grids, a modified interface operator is used that
reflects the geometry of the interface (see, e.g., \cite{almgren-iamr} for 
details of the multilevel cell-centered interface stencil).

In an algorithm without subcycling one can perform a composite
solve at every time step, as described in \cite{ricker_multigrid:2008},
to solve for $\phi$ on all levels.
Because the CASTRO algorithm uses subcycling in time, however,
we must use level solves at times when the solution is not defined
at all levels,   and then synchronize the solutions at different
levels as appropriate.
Even without changes in $\rho$ due to volume averaging and refluxing,
replacing a composite solve by separate level solves generates a 
mismatch in the normal gradient of $\phi$ at the boundary between
each level.  We correct these mismatches with a multilevel 
{\it correction solve}, which is a two-level composite solve for
a correction to $\phi.$  In addition to correcting the solutions
once the mismatch is detected, we add a correction term to later 
level solve solutions in order to minimize the magnitude of the
correction that will be needed.

%Both of the changes in $\rho$ that affect the monopole gravity solve 
%in multilevel simulations are relevant here; in addition, there
%is the issue of matching conditions at the coarse-fine interface that
%arises because we are approximating the composite solve by separate 
%level solves.  (Specifically, by doing separate single level solves
%we enforce Dirichlet matching conditions at the coarse-fine interface, but
%we do not enforce the Neumann matching conditions that would be satisfied in
%%a composite solve.) Thus, even if volume averaging and explicit
%refluxing do not change the density field, the solution to the Poisson
%equation as computed on the levels separately does not satisfy the multilevel
%operator.  This is discussed extensively in \cite{almgren-iamr}, for example; here we
%focus on how the gravity solves interact with the rest of the CASTRO algorithm.

\subsubsubsection{Multilevel Algorithm}

At the start of a calculation, we perform a composite solve from level 0 through 
$\ell_{\rm finest}$ to compute $\phi$ at all levels.   In addition, after every
regridding step that creates new grids at level $\ell+1$ and higher, 
a composite solve from level $\ell$ through $\ell_{\rm finest}$
is used to compute $\phi$ at those levels. 

Following an approach similar to that described in \cite{miniati-colella},
at the start and end of each level $\ell$ time step we perform a level
solve to compute $\phi^\ell.$ The difference between $\phi_\ell^{\rm comp}$ and 
$\phi^\ell$ at the start of the time step is stored in $\phi^{\ell,{\rm corr}}.$ 
This difference is added to $\phi^\ell$ at the beginning {\it and} end of this
level $\ell$ time step.   Thus $\phi^\ell + \phi^{\ell,{\rm corr}}$ is 
identical to $\phi_\ell^{\rm comp}$ at the start of the time step; at the
end of the time step it is an approximation to what the solution to the
composite solve would be.  In the event that the density
does not change over the course of the time step, the
effect of this lagged correction is to make  $\phi^\ell + \phi^{\ell,{\rm corr}}$
at the end of the time step identical to $\phi_\ell^{\rm comp}$ at that time,
thus there is no mismatch between levels to correct.
In general, when the density is not constant, the effect of the lagged
correction is to make the correction solve that follows the 
end of the time step much quicker.  We now describe the two-level correction step.
In the discussion below, we will refer to the two
levels involved in a correction solve as  the ``coarse'' and ``fine'' levels.

At the end of $r_\ell$ level $\ell+1$ time steps, when the level $\ell+1$
solution has reached the same point in time as the level $\ell$ solution,
and after the volume averaging and refluxing steps above have been performed,
we define two quantities on the coarse grid.  The first is 
the cell-centered quantity, $(\delta \rho)^{c},$ 
which carries the change in density at the
coarse level due only to refluxing.
%(\delta \rho)^{c} =  \rho^{c, \star} - {\overline{\rho}}^{c}  .
The second is the face-centered flux register, 
\begin{equation}
\delta F_\phi^\ell =  - A^c \frac{\partial \phi^c}{\partial n}
+ \sum A^f \frac{\partial \phi^f}{\partial n} ,
\end{equation}
which accounts for the mismatch in the normal gradient of 
$\phi$ at coarse-fine interfaces.
Here $A^c$ and $A^f$ represent area weighting factors on the
coarse and fine levels, respectively.
We define the composite residual, $R^{\rm comp},$
to be zero in all fine cells and in all coarse cells away
from the union of fine grids, and  
\begin{equation}
R^{\rm comp} = 4 \pi G (\delta \rho)^c - (\nabla \cdot \delta F_\phi ) |^c ,
\end{equation}
on all cells adjacent to the union of fine grids,
where $(\nabla \cdot) |^c $ refers to the discrete divergence at the coarse level,
where the only non-zero contribution comes from $\delta F_\phi$ on the coarse-fine interface.
We then solve
\begin{equation}
 L_{\ell,\ell+1}^{\rm comp} \; \delta \phi = R^{\rm comp}
\label{eq:gravsync}
\end{equation} 
and define the update to gravity at both levels,
\begin{equation}
\delta \gb = -\nabla (\delta \phi)  .
\end{equation}
This update is used to correct the gravitational source terms.  We define
the new-time state after volume averaging but before refluxing as 
$(\overline{\rho},\overline{\bf u}, \overline{\rho E},...),$
and the contributions to the solution on the coarse grid from refluxing as
$( (\delta \rho)^c, \delta (\rho {\bf u})^c, \delta (\rho E)^c, ...).$
Then we can define the sync sources for momentum on the coarse and fine levels, 
$S^{{\rm sync},c}_{\rho \ub}$, and $S^{{\rm sync},f}_{\rho \ub}$, respectively as follows:
\begin{eqnarray*}
S^{{\rm sync},c}_{\rho \ub} &=& \left(
        \overline{\rho}^c + (\delta \rho)^c \right) (\gb^{c,n+1} + \delta \gb^{c}) -
        \overline{\rho}^c                 \; \gb^{c,n+1}  \\
           &=& \left[( \delta \rho)^c \gb^{c,n+1} + 
                (\overline{\rho}^{c} + (\delta \rho)^c) \; \delta \gb^c)  \right ] \\
S^{{\rm sync},f}_{\rho \ub} &=& \overline{\rho}^{f} \; \delta \gb^f  .
\end{eqnarray*}
These momentum sources lead to the following energy sources:
\begin{eqnarray*}
S^{{\rm sync},c}_{\rho E} &=& S^{{\rm sync},c}_{\rho \ub} \cdot
         \left(   
    \overline{\ub}^{c} + \myhalf \; \Delta t_c \; S^{{\rm sync},c}_{\rho \ub} / \; \overline{\rho}^{c}
         \right) \\
S^{{\rm sync},f}_{\rho E} &=& S^{{\rm sync},f}_{\rho \ub} \cdot
         \left(
    \overline{\ub}^{f} + \myhalf \; \Delta t_f \; S^{{\rm sync},f}_{\rho \ub} / \; \overline{\rho}^{f}
         \right)
\end{eqnarray*}
The state at the coarse and fine levels is then updated using:
\begin{eqnarray*}
(\rho \ub)^{c,n+1} =  (\rho \ub)^{c} +  \delta (\rho \ub)^c + \myhalf \Delta t_c S^{{\rm sync},c}_{\rho \ub} &,&
\hspace{1em}
(\rho \ub)^{f,n+1} =  (\overline{\rho \ub})^{f} + \myhalf \Delta t_f S^{{\rm sync},f}_{\rho \ub}, \\
(\rho E)^{c,n+1} =  (\rho E)^{c} + \delta (\rho E)^c + \myhalf \Delta t_c S^{{\rm sync},c}_{\rho E} &,&
\hspace{1em}
(\rho E)^{f,n+1} =  (\overline{\rho E})^{f} + \myhalf \Delta t_f S^{{\rm sync},f}_{\rho E}  .
\end{eqnarray*}
(The factor of $\myhalf$ follows from the time-centering of the sources.)

To complete the correction step,
\begin{itemize}
\item we add $\delta \phi$ directly to 
$\phi^{\ell}$ and $\phi^{\ell+1}$ and interpolate $\delta \phi$ to any finer
levels and add it to the current $\phi$ at those levels.  We note that at this
point $\phi$ at levels $\ell$ and $\ell+1$ is identical to the solution that would 
have been computed using a two-level composite solve with the current values of density.   
Thus the new, corrected, $\phi$ at each level plays the role of $\phi^{\rm comp}$ in the next time step.

\item if level $\ell > 0$, we transmit the
effect of this change in $\phi$ to the coarser levels by updating the flux register between
level $\ell$ and level $\ell-1.$ In particular, we set
\begin{equation}
\delta {F_\phi}^{\ell-1} = \delta {F_\phi}^{\ell-1} 
+ \sum A^c \frac{\partial (\delta \phi)^{c-f}}{\partial n}  .
\end{equation}
\end{itemize}

\subsubsubsection{Performance Issues}
%In some simulations the use of $\phi^{\ell,\rm corr}$ provides a negligible improvement 
%to the accuracy of the overall solution.   In this case
%we have a {\it no\_composite} flag that eliminates all non-correction multilevel solves, 
%and consequently the lagged correction as well.    We still perform the two-level
%correction solve as described above.  We also have a {\it no\_sync} option that
%%eliminates all of the two-level correction solves as well; this is useful
%for the case where a monopole approximation is not appropriate yet the coarse-fine
%interface is sufficiently far from the dominant massive structure that the 
%synchronization step provides negligible improvement as well.

The multilevel algorithm is not as computationally expensive as it
might appear.  Because multigrid is an iterative solver, 
the cost of each solve is proportional to the number of V-cycles, 
which is a function of the desired reduction in residual.   We can reduce the
number of V-cycles needed in two ways.  First, we can supply a good initial guess for
the solution; second, we can lower the desired reduction in residual.

In the case of level solves, we always use $\phi$ from a previous level
solve, when available, as a guess in the current level solve.  Thus, even in
a single-level calculation, $\phi$ from the beginning of the time step is used
as a guess in the level solve at the end of the time step.  If no regridding
occurs, then $\phi$ at the end of one time step can be used as a guess for
the level solve at the start of the next time step.  The extent to which $\rho$
changes in a time step dictates the extent to which a new computation of gravity
is needed, but this also dictates the cost of the update.

Similarly, there is no point in solving for $\delta \phi$ to greater accuracy than
we solve for $\phi.$ When we do the correction solve for $\delta \phi$, we require only that the 
residual be reduced to the magnitude of the final residual from the level solve, 
{\it not} that we reduce the correction residual by the same factor.
Thus, if the right-hand-side for the correction solve is already small, 
the cost of the correction solve will be significantly
less than that of the initial level solve.

\section{Software Design and Parallel Performance}

\subsection{Overview}

CASTRO is implemented within the BoxLib framework,
a hybrid \Cplusplus \slash Fortran90 software system that provides
support for the development of parallel structured-grid AMR applications.
The basic parallelization strategy uses a hierarchical programming approach
for multicore architectures based on both MPI and OpenMP.   
In the pure-MPI instantiation, at least one grid at each level is 
distributed to each core, and each core communicates
with every other core using only MPI.  In the hybrid approach, where 
on each socket there are $n$ cores which all access the same memory, 
we can instead have one larger grid per socket,
with the work associated with that grid distributed among the $n$ cores
using OpenMP.    

In BoxLib, memory management, flow control, parallel communications and I/O are
expressed in the \Cplusplus portions of the program.  The numerically intensive portions of the 
computation, including the multigrid solvers, are handled in Fortran90.
The fundamental parallel abstraction in both the \Cplusplus  and the Fortran90 is the 
MultiFab, which holds the data on the union of grids at a level.  A MultiFab is
composed of FAB's; each FAB is an array of data on a single grid. 
During each MultiFab operation the FAB's composing that MultiFab 
are distributed among the cores.
MultiFab's at each level of refinement are distributed
independently.  The software supports two data distribution schemes, 
as well as a dynamic switching scheme that decides which approach to use 
based on the number of grids at a level and the number of processors.
The first scheme is based on a heuristic knapsack algorithm as
described in \citet{crutchfield:1991} and in \citet{rendleman-hyper}. 
The second is based on the use of a Morton-ordering space-filling curve.
%MultiFab operations are performed with an \emph{owner computes} rule
%with each processor operating independently on its local
%data.  For operations that require data owned by other processors, the
%MultiFab operations are preceded by a data exchange between processors.

Each processor contains {\it meta-data} that
is needed to fully
specify the geometry and processor assignments of the MultiFab's.  At
a minimum, this requires the storage of an array of
boxes specifying the index space
region for each AMR level of refinement.
%The meta-data can thus be used to dynamically evaluate the necessary
%communication patterns for sharing data amongst processors enabling us
%to optimize communications patterns within the algorithm.  
One of the advantages of computing with fewer, larger grids in the hybrid
OpenMP--MPI approach is that the size of the meta-data is substantially
reduced.

\subsection{Parallel Output}

Data for checkpoints and analysis are written in a self-describing
format that consists of a directory for each time step written.
Checkpoint directories contain all necessary data to restart the
calculation from that time step.  Plotfile directories contain data
for postprocessing, visualization, and analytics, which can be read
using \Code{amrvis}, a customized visualization package developed at
LBNL for visualizing data on AMR grids, or VisIt \citep{visit}.  
Within each checkpoint or plotfile
directory is an ASCII header file and subdirectories for each AMR
level.  The header describes the AMR hierarchy, including number of
levels, the grid boxes at each level, the problem size, refinement
ratio between levels, step time, etc.  Within each level directory are
the MultiFab files for each AMR level.  Checkpoint and plotfile
directories are written at user-specified intervals.

%but differ in the type of data written.
%can be written at different step intervals.  For example,
%the calculation can write Checkpoints every 100 steps
%and Plotfiles every 20 steps.  These intervals are set
%at runtime.

For output, each processor writes its own data to the appropriate
MultiFab files.  The output streams are coordinated to only allow one
processor to write to a file at one time and to try to maintain
maximum performance by keeping the number of open data streams, which
is set at run time, equal to the number of files being written.  Data
files typically contain data from multiple processors, so each
processor writes data from its associated grid(s) to one file, then 
another processor can write data from its associated grid(s) to that 
file.  A designated I/O Processor writes the header files
and coordinates which processors are allowed to write to which files
and when.  The only communication between processors is for signaling
when processors can start writing and for the exchange of header
information.  We also use the \Cplusplus \Code{setbuf} function for good single 
file performance.  While I/O performance even during a single run can
be erratic, recent timings on the Franklin machine (XT4) at NERSC 
indicate that CASTRO's I/O performance, when run with a single level 
composed of multiple uniformly-sized grids, matches some of the 
top results for the N5 IOR benchmark (roughly 13GB/s) \citep{nersc_io}.  
For more realistic simulations with multiple grids at multiple levels,
CASTRO is able to write data at approximately 5 GB/s sustained,
over half of the average I/O benchmark reported speed.

\subsection{Parallel Restart}

Restarting a calculation can present some difficult issues
for reading data efficiently.  In the worst case,
all processors would need data from all files.
If multiple processors try to read from the same file at
the same time, performance problems can result,
with extreme cases causing file system thrashing.
Since the number of files is generally not equal to the
number of processors and each processor may need data
from multiple files, input during restart is coordinated to
efficiently read the data.  Each data file is only opened by one
processor at a time.
The IOProcessor creates a database for mapping files
to processors, coordinates the read queues, and interleaves
reading its own data.  Each processor reads all data it needs
from the file it currently has open.
%in \Code{lseek} order.
The code tries to maintain the number of input streams to be
equal to the number of files at all times.

Checkpoint and plotfiles are portable to machines with a
different byte ordering and precision from the machine that wrote the files.
Byte order and precision translations are done automatically,
if required, when the data is read.

\subsection{Parallel Performance}

\begin{figure}[t]
\centering
\includegraphics[width=6.75in]{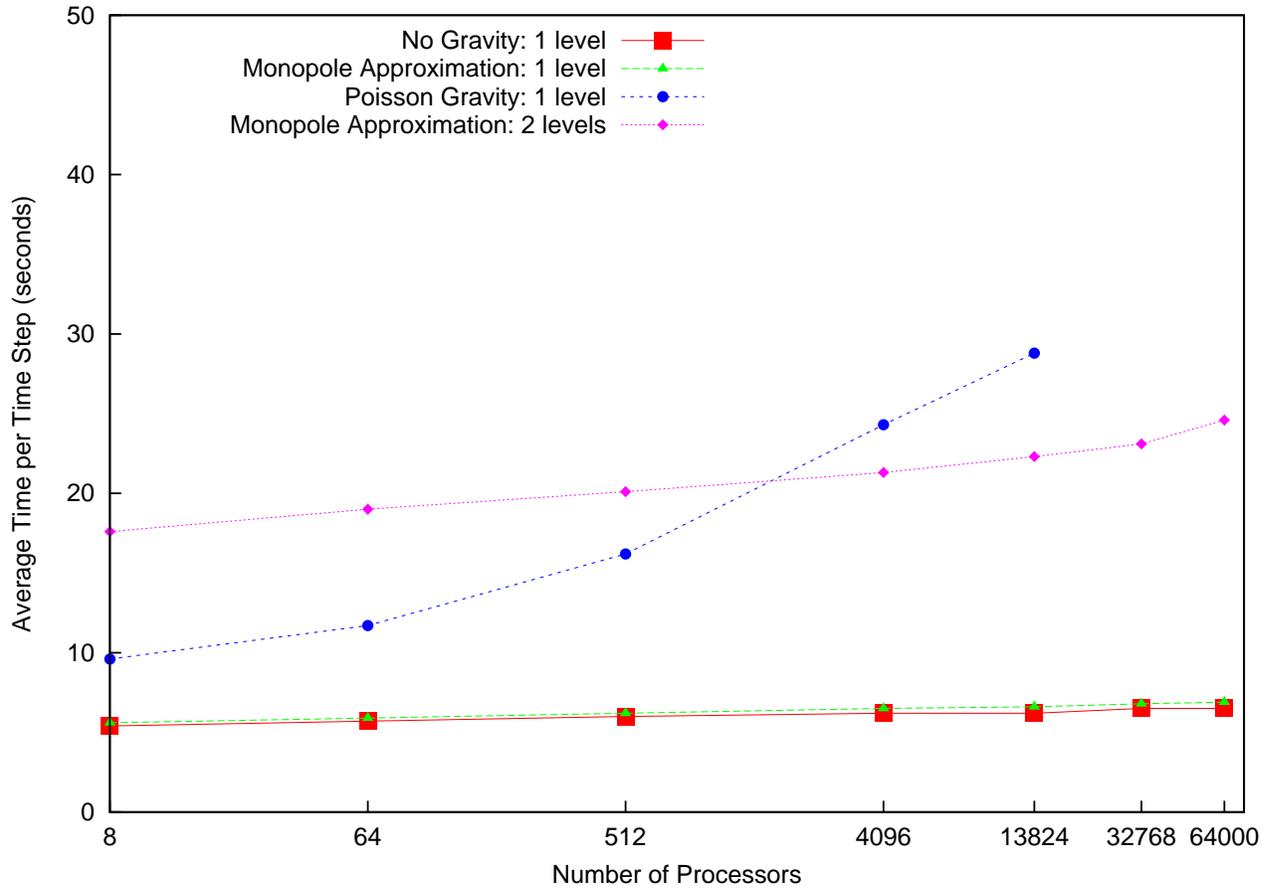}
\caption{\label{fig:scaling} Weak scaling behavior of the CASTRO code on  the jaguarpf machine at the
OLCF.  For the two-level simulation, the number of cells that are advanced in a time step
increases by a factor of three because of subcycling.
To quantify the overall performance, we note that for the 64,000 processor case without
gravity, the time for a single core to advance one cell for one time step is 24.8~$\mu$s.}
\end{figure}

In Figure~\ref{fig:scaling} we show the scaling behavior of the CASTRO code, 
using only MPI-based parallelism, on the
jaguarpf machine at the Oak Ridge Leadership Computing Facility (OLCF).   
A weak scaling study was performed, so that for each run
there was exactly one $64^3$ grid per processor.  We ran the code with gravity
turned off, with the monopole approximation to gravity, and with the Poisson solve for
gravity.   The monopole approximation to gravity adds very little
to the run time of the code; with and without the monopole approximation the
code scales excellently from 8 to 64,000 processors.   
For the 64,000 processor case without gravity, the time for a single core to 
advance one cell for one time step is 24.8~$\mu$s.

Good scaling of linear solves is known to be much more difficult to achieve; 
we report relatively good scaling up to only 13,824 processors in the
pure-MPI approach.   An early strong scaling study contrasting the pure-MPI
and the hybrid-MPI-OpenMP approaches for a $768^3$ domain shows that
one can achieve at least a factor of 3 improvement in linear solver time
by using the hybrid approach at large numbers of processors.  Improving the 
performance of the linear solves on the new multicore architectures is an area 
of active research; more extensive development and testing is underway.

We also ran a scaling study with a single level of local refinement using the monopole
gravity approximation.  In this MPI-only study, there is one $64^3$ grid at each level 
for each processor.  Because of subcycling in time, a coarse time step consists of a single step
on the coarse grid and two steps on the fine grid. Thus, we would expect that 
the time to advance the multilevel solution by one coarse time step would 
be a factor of three greater than the time to advance the single-level coarse
solution by one coarse time step, plus any additional overhead associated with AMR.  
From the data in the figure we conclude that AMR
%modest increase in time per step in addition to 3 times the single level simulation 
%time over the range of processors being considered, suggesting that AMR 
introduces a modest overhead, ranging from approximately 5\% for the 8 processor case to
19\% for the 64,000 processor case.  By contrast, advancing a 
single-level calculation at the finer resolution by the same total time, i.e., 
two fine time steps, would require a factor of 16 more resources than advancing the 
coarse single-level solution.

\section{Test Problems}

In this section we present a series of calculations demonstrating the
behavior of the hydrodynamics, self-gravity, and reaction components
of CASTRO.  The first set contains three one-dimensional shock tube
problems, including Sod's problem, a double rarefaction problem, and a
strong shock problem.  We follow this with Sedov-Taylor blast waves
computed in 1D spherical coordinates, 2D cylindrical and Cartesian
coordinates, and 3D Cartesian coordinates.  Our final
pure-hydrodynamics test is a 2D Rayleigh-Taylor instability.  We use
this problem to contrast the differences in the flow found using
dimensionally split and unsplit methods with piecewise linear, PPM 
with the old limiters, and PPM with the new limiters. 

We then present two examples that test the interaction of the
self-gravity solvers with the hydrodynamics in 3D Cartesian
coordinates.  In the first case a star is initialized in hydrostatic
equilibrium and we monitor the maximum velocities that develop; in the
second, the homologous dust collapse test problem, a uniform-density
sphere is initialized at a constant low pressure, and collapses under
its own self-gravity.  These tests more closely examine the 3D
spherical behavior we expect to be present in simulations of Type Ia
and Type II supernovae.

We perform a test of the coupling of the hydrodynamics to reactions.
This test consists of a set of buoyant reacting bubbles in a
stratified stellar atmosphere.  We compare the CASTRO results to those
of the FLASH code.

Finally, we note that a previous comparison of CASTRO to our low Mach
number hydrodynamics code, MAESTRO, can be found in \citet{multilevel}.
In that test, we took a 1-d spherical, self-gravitating stellar model and
watched it hydrostatically adjust as we dumped energy into the center
of the star.  The resulting temperature, pressure, and density profiles
agreed very well between the two codes.

\subsection{Shock Tube Problems}  
To test the behavior of the hydrodynamics solver, we run several
different 1D shock tube problems.  The setup for these problems
consists of a left and right state, with the interface in the center
of the domain.  All calculations use a gamma-law equation of state
with $\gamma = 1.4.$ We show results from each problem run using 1D
Cartesian coordinates, but we have verified that the results are
identical when each problem is run in 2D or 3D Cartesian coordinates
and the interface is normal to a coordinate axis.
%In multiple dimensions, the transverse directions are initialized uniformly.  
The length of the domain is always taken as 1.0, with the interface in the center.
We use a base grid of 32 cells, with two additional levels of factor
2 refinement, for an effective resolution of 128 cells.   
The refinement criteria are based on gradients of density and velocity.
In the case of the double rarefaction we also present results from runs with
two levels of factor 4 refinement (effective resolution of 512 cells) and 
three levels of factor 4 refinement (effective resolution of 2048 cells).
In each case, analytic solutions are found using the exact Riemann solver from \citet{Toro}.
All calculations are run with the new PPM limiters and a CFL number of 0.9.
For each problem we show density, pressure, velocity, and internal energy.

\subsubsection{Sod's Problem}

\begin{figure}[t]
\centering
\includegraphics[width=6.75in]{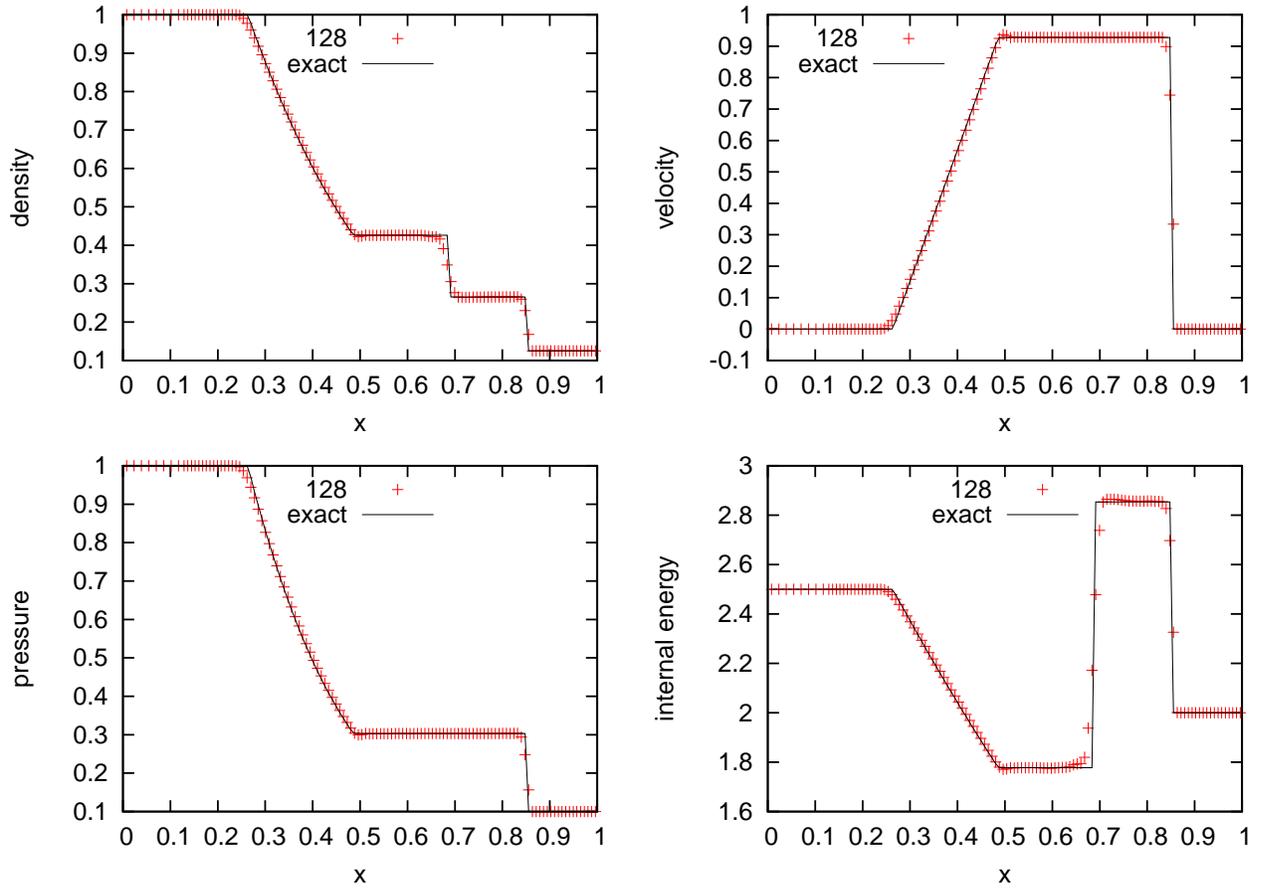}
\caption{\label{fig:sod} Adaptive CASTRO solution vs.\ analytic solution for Sod's problem
                         run in 1D at an effective resolution of 128 cells.}
\end{figure}

The Sod problem \citep{sod:1978} is a simple shock tube problem that
exhibits a shock, contact discontinuity, and a rarefaction wave.
The non-dimensionalized initial conditions are:
\begin{equation}
\begin{array}{l}
\rho_L = 1 \\
u_L = 0 \\
p_L = 1
\end{array} 
\qquad
\begin{array}{l}
\rho_R = 0.125 \\
u_R = 0 \\
p_R = 0.1
\end{array}
\end{equation} 
This results in a rightward moving shock 
and contact discontinuity, and a leftward moving rarefaction wave.
Figure~\ref{fig:sod} shows the resulting pressure, density, velocity,
and internal energy at $t= 0.2$~s.  We see excellent agreement with
the exact solution.

\subsubsection{Double Rarefaction}

\begin{figure}[t]
\centering
\includegraphics[width=6.75in]{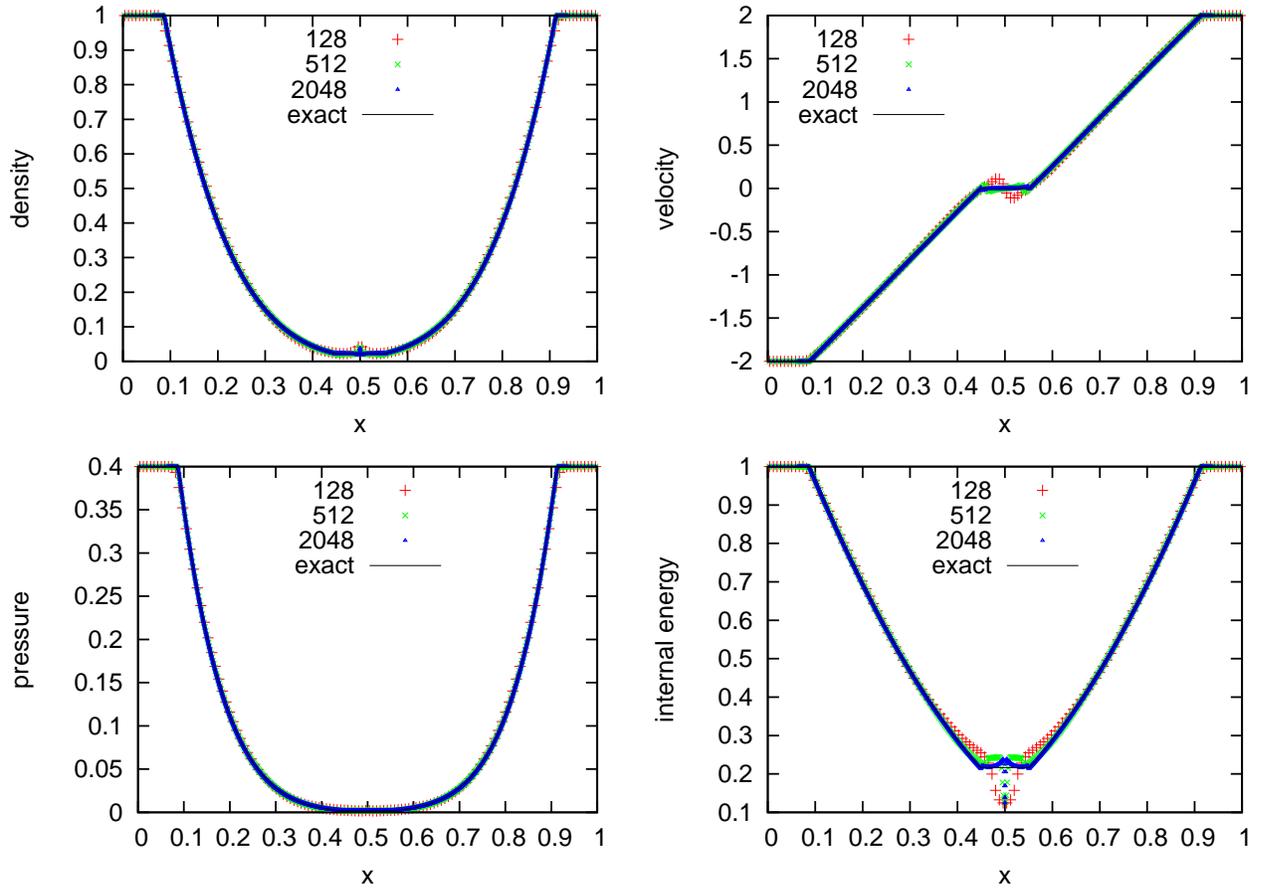}
\caption{\label{fig:test2} Adaptive CASTRO solutions vs.\ analytic solution for 
  the double rarefaction problem run in 1D at effective resolutions of 128, 512 and 
  2048 cells.}
\end{figure}

The double rarefaction problem tests the behavior of the hydrodynamics
algorithm in regions where a vacuum is created.  We run the problem as
described in \citet{Toro}.  The non-dimensionalized initial conditions are:
\begin{equation}
\begin{array}{l}
\rho_L = 1 \\
u_L = -2 \\
p_L = 0.4
\end{array}
\qquad
\begin{array}{l}
\rho_R = 1 \\
u_R = 2 \\
p_R = 0.4
\end{array}
\end{equation}
This results in two rarefaction waves propagating in opposite directions away 
from the center.  As a result, matter is evacuated from the center, leaving behind
a vacuum.  Figure~\ref{fig:test2} shows the CASTRO solutions at $t = 0.15$~s.
%The CASTRO solutions are all run
%with a base grid of 32 cells; the effective 128-cell resolution uses
%two levels of factor two refinement, the effective 512-resolution uses
%two levels of factor four refinement, and the effective 2048-resolution
%case uses three levels of factor four refinement.   In all cases the
%CFL number is 0.9. 
The agreement with the exact solution is excellent at the 128-cell resolution for
density, pressure and velocity; the internal energy is more sensitive,
but clearly converges to the analytic solution except at the center line.
This is a very common pathology for
this problem, since the internal energy, $e$, is derived from
equation~(\ref{eq:gamma-law}) using values of $p$ and $\rho$ which are both
approaching zero in the center of the domain \citep{Toro}.  
%\MarginPar{the energy drops at the center, but it used to be above
%the exact solution---this is a CFL or init\_shrink effect---do we want to comment?}

\subsubsection{Strong Shock}

\begin{figure}[t]
\centering
\includegraphics[width=6.0in]{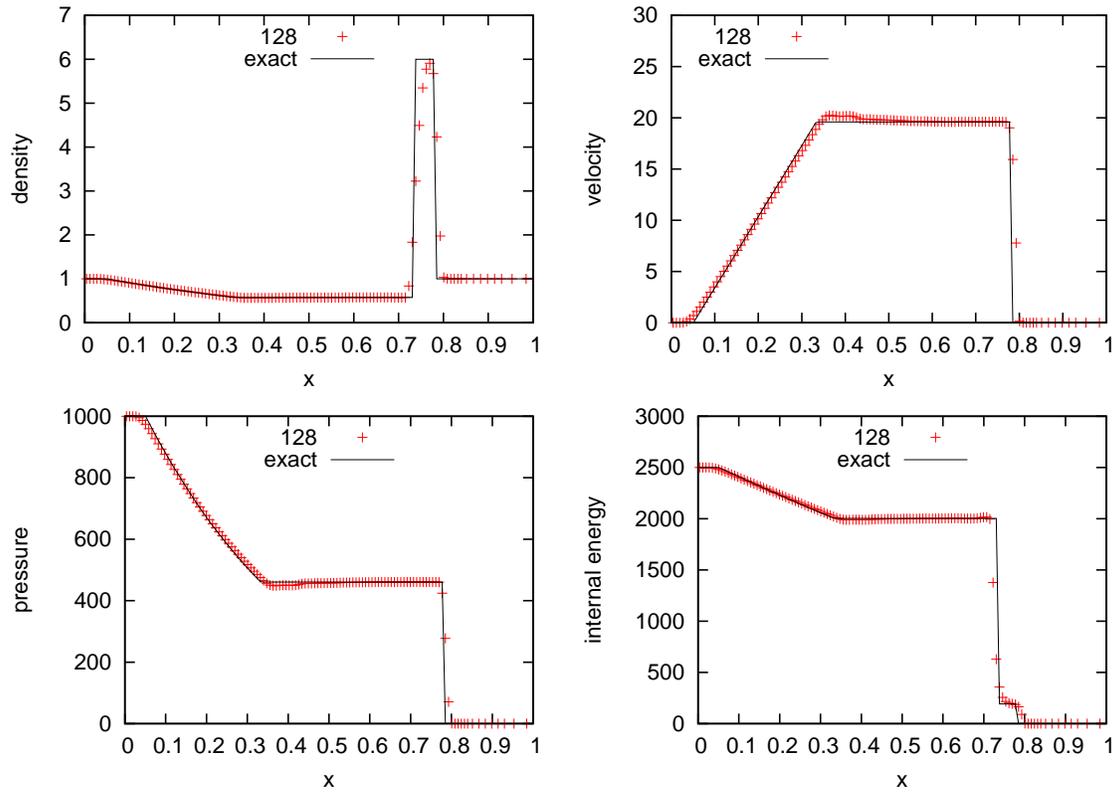}
\caption{\label{fig:test3} Adaptive CASTRO solution vs.\ analytic solution for the strong
                           shock problem run in 1D at an effective resolution of 128 cells.}
\end{figure}

The final shock-tube problem we try is a strong shock.  We initialize
the problem as described in \citet{Toro}.  The initial conditions are:
\begin{equation}
\begin{array}{l}
\rho_L = 1 \\
u_L = 0 \\
p_L = 1000
\end{array}
\qquad
\begin{array}{l}
\rho_R = 1 \\
u_R = 0 \\
p_R = 0.01
\end{array}
\end{equation}
The initial pressure jump of six orders of magnitude results in 
a strong rightward moving shock.  This large dynamic range can cause 
trouble for some hydrodynamics solvers.
The shock is followed very closely by a contact
discontinuity.  A leftward moving rarefaction is also present.
Figure~\ref{fig:test3} shows the CASTRO results at $t = 0.012$~s.
We see good agreement between the CASTRO results and the exact
solution.  
%The only real disagreement is see in the density field
%and the contact discontinuity, owing to the severity of the shock.
%Increased resolution would improve the solution.  

\subsection{Sedov}

Another standard hydrodynamics test is the Sedov-Taylor blast wave.
The problem setup is very simple: a large amount of energy is
deposited into the center of a uniform domain.  This drives a blast
wave (spherical or cylindrical, depending on the domain geometry).  An
analytic solution is provided by \citet{sedov:1959}.  We use a
publicly available code described by \citet{timmes_sedov_code} to
generate the exact solutions.

The Sedov explosion can test the geometrical factors in the
hydrodynamics scheme.  A cylindrical blast wave (e.g.\ a point
explosion in a 2D plane) can be modeled in 2D Cartesian coordinates.
A spherical blast wave can be modeled in 1D spherical, 2D
axisymmetric (cylindrical $r$-$z$), or 3D Cartesian coordinates.

In the Sedov problem, the explosion energy, $\Eexpl$
(in units of energy, not energy/mass or energy/volume), is deposited into
a single point, in a medium of uniform ambient density,
$\rho_\mathrm{ambient}$, and pressure, $p_\mathrm{ambient}$.
Initializing the problem can be difficult because the small volume is
typically only one cell in extent,  which can lead to grid imprinting in
the solution.  A standard approach (see for example \citealt{flash,omang:2006}
and the references therein) is to convert the explosion energy into a
pressure contained within a certain volume, $V_\mathrm{init}$, of
radius $r_\mathrm{init}$ as
\begin{equation}
p_\mathrm{init} = \frac{(\gamma - 1) \Eexpl}{V_\mathrm{init}}.
\end{equation}
This pressure is then initialized to $p_\mathrm{init}$ 
in all of the cells where $r < r_\mathrm{init}$.  
We use the gamma-law equation of state with $\gamma = 1.4$.

To further minimize any grid effects, we do subsampling
in each cell: each cell is divided it into $N_\mathrm{sub}$ subcells in each
coordinate direction, each subcell is initialized independently, and
then the subcells are averaged together (using volume weighting for
spherical or cylindrical coordinates) to determine the
initial state of the full cell.

For these runs, we use $\rho_\mathrm{ambient} = 1~\gcc$,
$p_\mathrm{ambient} = 10^{-5}~\presu$, $\Eexpl = 1~\mathrm{erg}$, $r_\mathrm{init}
= 0.01~\mathrm{cm}$, and $N_\mathrm{sub} = 10$.  A base grid with $\Delta x =
0.03125$~cm is used with three levels of factor 2 refinement.  For most
geometries, we model the explosion in a domain ranging from 0 to
1~cm in each coordinate direction.  In this case, the base grid would have
32 cells in each coordinate direction and the finest mesh
would correspond to 256 cells in each coordinate direction.  For the 2D
axisymmetric case, we model only one quadrant,
%\MarginPar{for the 3D run, do we want to show the density field with the grids overlaid?}  
and the domain ranges from 0 to 0.5~cm.  All calculations were run
with a CFL number of 0.5, and the initial time step was shrunk by a
factor of 100 to allow the point explosion to develop.  We refine on
regions where $\rho > 3~\gcc$, $\nabla \rho >
0.01~\gcc~\mathrm{cm}^{-1}$, $p > 3~\presu$, or $\nabla p >
0.01~\presu~\mathrm{cm}^{-1}$. 

\begin{figure}[t]
\centering
\includegraphics[width=6.0in]{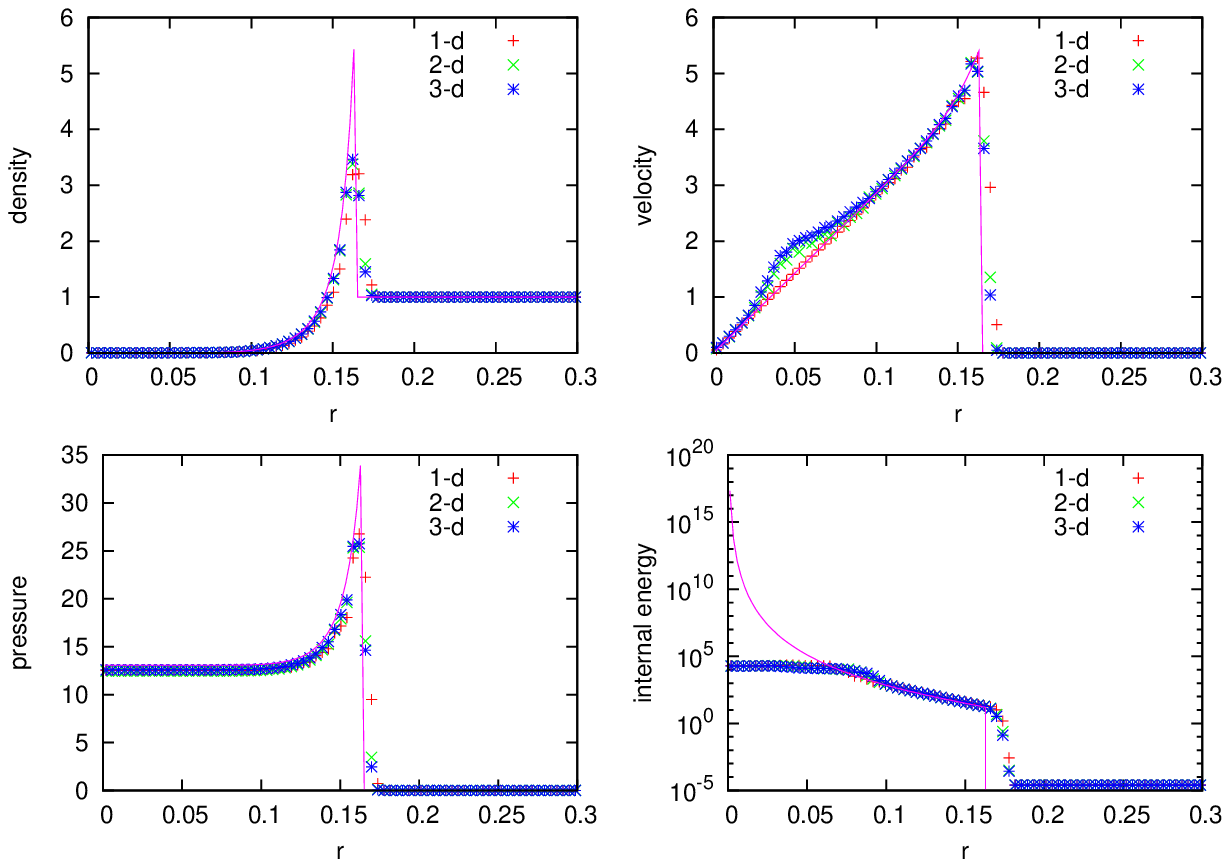}
\caption{\label{fig:sedov_sph_3lev} CASTRO solution at $t=0.01$s 
  for the spherical Sedov blast wave problem
  run in 1D spherical, 2D axisymmetric, and 3D Cartesian coordinates.
  This was run with a base grid with $\Delta x = 0.03125$~cm and 3 levels of factor 2 refinement
  for an effective resolution of  $\Delta x = .00390625$~cm.}
\end{figure}

\begin{figure}[t]
\centering
\includegraphics[width=6.0in]{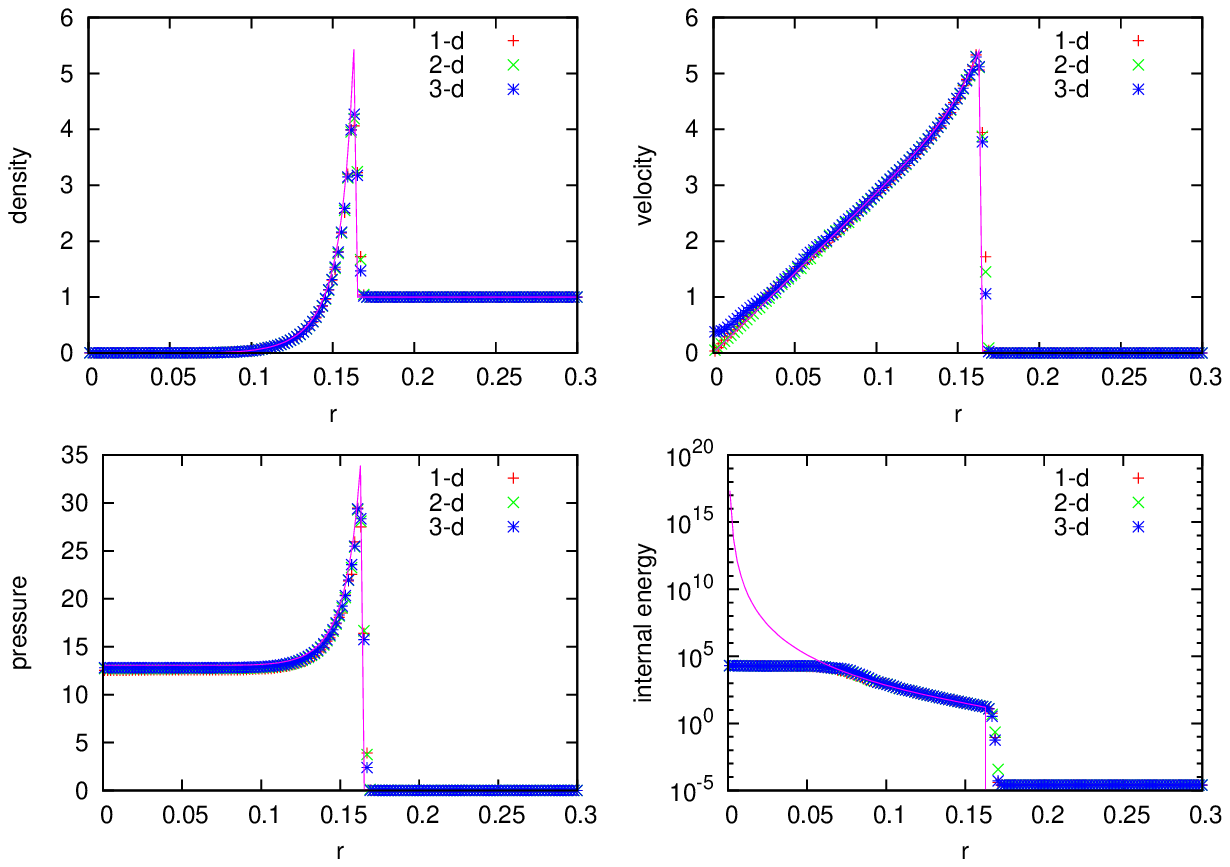}
\caption{\label{fig:sedov_sph_4lev} CASTRO solution at $t=0.01$s 
  for the spherical Sedov blast wave problem
  run in 1D spherical, 2D axisymmetric, and 3D Cartesian coordinates.
  This was run with a base grid with $\Delta x = 0.03125$~cm and 4 levels of factor 2 refinement
  for an effective resolution of  $\Delta x = .001953125$~cm.}
\end{figure}

\begin{figure}[t]
\centering
\includegraphics[width=6.0in]{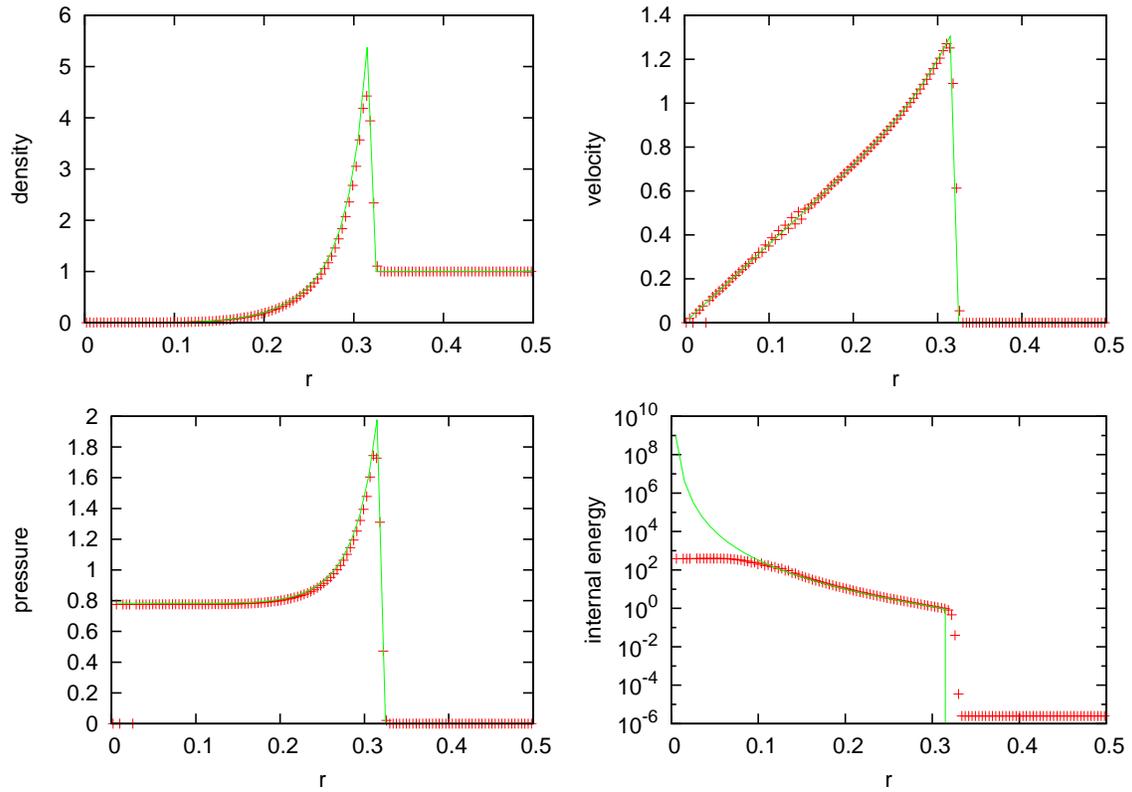}
\caption{\label{fig:sedov_cyl} CASTRO solution at $t=0.1$s for the cylindrical Sedov blast wave problem
  run in 2D Cartesian coordinates.  This was run with a base grid with $\Delta x = 0.03125$~cm 
  and 3 levels of factor 2 refinement for an effective resolution of  $\Delta x = .00390625$~cm.}
\end{figure}

Figure~\ref{fig:sedov_sph_3lev} shows the CASTRO solution to a 
spherical Sedov explosion at time $t=0.01$s, run in 1D spherical, 2D cylindrical, and
3D Cartesian coordinates. For the 2D and 3D solutions, we
compute the radial profile by mapping each cell into its
corresponding radial bin and averaging.  The radial bin width was
picked to match the width of a cell at the finest level of refinement
in the CASTRO solution.  The density, velocity, and pressure plots
match the exact solution well.  As with the double rarefaction problem, 
the internal energy is again the most difficult quantity to match
due to the vacuum region created at the origin.
Figure~\ref{fig:sedov_sph_4lev} shows the same set of calculations run
with 4 levels of factor 2 refinement.  Here the agreement is even better. 
%\MarginPar{there is a departure of the 3D velocity at the origin.  Also,
%do we want to show both resolutions?}
Figure~\ref{fig:sedov_cyl} shows the CASTRO solution at time $t=0.1$s to a cylindrical
Sedov explosion, run in 2D Cartesian coordinates.

\subsection{Rayleigh-Taylor}

The Rayleigh-Taylor instability results when a dense fluid is placed
over a less-dense fluid in a gravitational field
\citep{taylor:1950,layzer:1955,sharp:1984}.  The interface is unstable
and a small perturbation will result in the growth a buoyant uprising
bubbles and dense, falling spikes of fluid.  This instability provides
a mechanism for mixing in many astrophysical systems.  Despite its
seemingly simplistic nature, only the linear growth regime is
understood analytically (see for example \citealt{chandrasekharbook}).
In the non-linear regime, Rayleigh-Taylor instability calculations are
often used as a means of code validation \citep{RTalpha}.

For our purposes, the R-T instability provides a good basis to compare
different choices of the advection algorithm.  We model a single-mode
Rayleigh-Taylor instability---a perturbation consisting of a single
wavelength that disturbs the initial interface.  Short-wavelength
perturbations have a faster growth rate than long-wavelength
perturbations, so grid effects can easily drive the instability on
smaller scales than our initial perturbation.  No viscous terms are
explicitly modeled.

We choose the density of the dense fluid to be $\rho_2 = 2~\gcc$ and
the light fluid is $\rho_1 = 1~\gcc$.  The gravitational acceleration
is taken to be $g = -1~\cmss$ in the vertical direction.  The
gamma-law equation of state is used with $\gamma = 1.4$.  Our domain
has a width of $L_x = 0.5$~cm and a height of $L_y = 1$~cm.  The
initial interface separating the high and low density fluid is
centered vertically at $L_y/2$, with the density in the top half taken
to be $\rho_2$ and the density in the lower half $\rho_1$.  Since $g$
and $\rho_1$, $\rho_2$ are constant, we can analytically integrate the
equation of hydrostatic equilibrium to get the pressure in both the
high and low-density regions of the domain:
\begin{equation}
p(y) = \left \{ \begin{array}{ll}
       p_\mathrm{base} + \rho_1 g y & y < L_y / 2 \\
       p_\mathrm{base} + \rho_1 g L_y/2 + \rho_2 g (y - L_y/2) \quad & y > L_y/2 
      \end{array}
       \right .
\end{equation}
where $y$ is the vertical coordinate, and $p_\mathrm{base}$ is the
pressure at the base of the domain.  We take $p_\mathrm{base} =
5~\presu$.

To initiate the instability, the interface is perturbed by slightly
shifting the density, keeping the interface centered vertically in the
domain.  We define the perturbed interface height, $\psi$, to be a function of
position in the $x$-direction as
\begin{equation}
\psi(x) = \frac{A}{2} \left [ \cos \left(\frac{2\pi x}{L_x} \right) + 
                               \cos \left(\frac{2\pi (L_x - x)}{L_x} \right )
                      \right ] 
           + \frac{L_y}{2}
\end{equation}
with the amplitude, $A = 0.01$~cm.  We note that the cosine part of 
%\MarginPar{note: by this construction, the initial state is not in HSE}
the perturbation is done symmetrically, to prevent roundoff error
from introducing an asymmetry in the flow.  The density is then perturbed
as:
\begin{equation}
\rho(x,y) = \rho_1 + \frac{\rho_2 - \rho_1}{2} 
           \left [ 1 + \tanh \left ( \frac{y - \psi(x)}{h} \right ) \right ]
\end{equation}
The $\tanh$ profile provides a slight smearing of the initial interface,
over a smoothing length $h$.  We take $h = 0.005$~cm. 

\begin{figure}[t]
\centering
\includegraphics[width=4.5in]{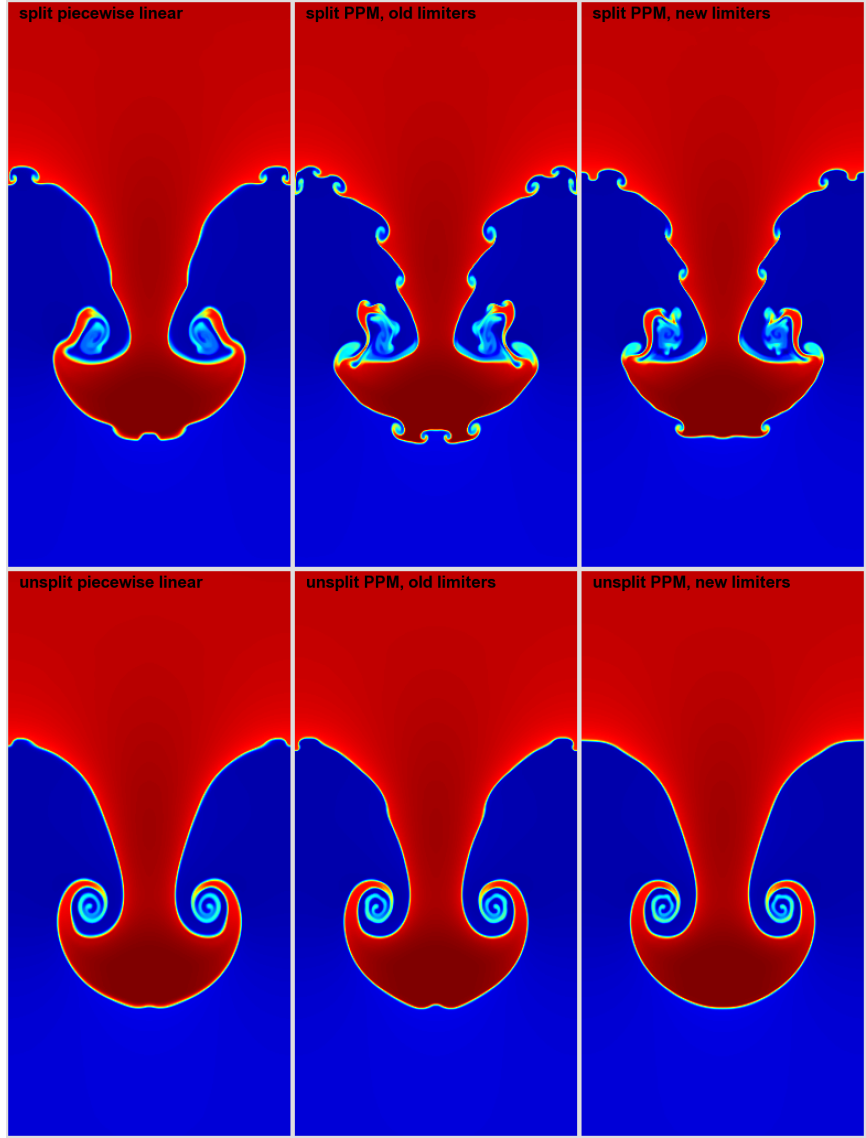}
\caption{\label{fig:RT} Density in a single-mode Rayleigh-Taylor
  simulation for a variety of advection schemes.  Dimensionally-split
  method results are shown on the top row; unsplit method results are
  shown on the bottom row.  We see that the unsplit methods do better
  at suppressing the growth of high-wavenumber instabilities resulting
  from grid effects.}
\end{figure}

In Figure \ref{fig:RT}, we present simulation results for the Rayleigh-Taylor
problem at $t = 2.5$s for several different variants of the hydrodynamics.
All calculations were run with $256 \times 512$ grid cells.
In the bottom right image we show the results obtained using the unsplit
PPM with the new limiter used in CASTRO.  The left
and middle images on the bottom row are results using the unsplit piecewise
linear method and unsplit PPM with limiters as in \cite{ppmunsplit}, respectively.
The results with all three methods are reasonably
good; however, the piecewise linear and original PPM limiter both exhibit mild
anomalies at the tip of both the bubble and the spike. 

In the upper row, we present results for the Rayleigh-Taylor problem using operator-split
analogs of the unsplit methods.
The details of the algorithms such as limiters, Riemann solver, etc. are the same as in the
unsplit methods; the only difference is the use of operator splitting. We note that all
three of the operator-split methods produce spurious secondary instabilities. 
This behavior is a direct result of the operator-split approach.  Physically, for these
low Mach number flows, the density field is advected by a nearly incompressible flow field,
and remains essentially unchanged along Lagrangian trajectories.  However, in regions
where there is significant variation in the local strain rate, an operator-split
integration approach alternately compresses and expands the fluid between subsequent sweeps.
This alternating compression / expansion provides the seed for the anomalies observed
with operator-split methods.

We note that both the CPU time and the memory usage are roughly a factor
of two larger for the unsplit algorithm than for the split algorithm
in this two-dimensional implementation.   For a pure hydrodynamics problem with gamma-law
equation of state this factor is nontrivial; for a simulation that uses
the full self-gravity solver, a realistic reaction network,
a costly equation of state, or significant additional physics, 
the additional cost of the hydrodynamic solver may be negligible.

In 3D one might expect the ratio of CPU time for the unsplit algorithm
relative to the split algorithm to be be even larger than in 2D
because of the additional Riemann solves required to construct the 
transverse terms. 
However, this effect is counterbalanced by the need to advance ghost cells in the
split algorithm to provide boundary conditions for subsequent sweeps.
Consequently, we observe an increase in CPU time that is slightly less than the factor
of two observed in 2D. 
The 3D implementation of the 
unsplit algorithm in CASTRO uses a strip-mining approach that only stores extra
data on a few planes at a time, so we see an increase of less than 10\% in the
memory required for the unsplit integrator compared to the split integrator
in 3D.
%In 3D we would expect the ratio of CPU time for the unsplit algorithm
%relative to the split algorithm to be closer to four than two 
%because of the additional Riemann solves required to construct the 
%transverse terms.   However, because the 3D implementation of the 
%unsplit algorithm in CASTRO operates on only a 
%subregion of the domain at a time (even if only a single grid were present),
%the ratio of memory usage is roughly the same in 3D as in 2D.

\subsection{Stationary Star Gravity}

A challenging problem for a hydrodynamics code is to keep a star in
hydrostatic equilibrium.  Because of the different treatment of the
pressure, density, and gravitational acceleration by the hydrodynamics
algorithm, small motions can be driven by the inexact cancellation of
$\nabla p$ and $\rho \gb$.  This is further exaggerated by modeling a
spherical star on a 3D Cartesian grid.  Here we test the ability of
CASTRO to maintain hydrostatic equilibrium for a spherical,
self-gravitating star.

Our initial model is a nearly-Chandrasekhar mass, carbon-oxygen white dwarf,
which is generated by specifying a core density
($2.6\times 10^9~\gcc$), temperature ($6\times 10^8$~K), and a uniform
composition ($X(^{12}\mathrm{C}) = 0.3, X(^{16}\mathrm{O}) = 0.7$) and
integrating the equation of hydrostatic equilibrium outward while
constraining the specific entropy, $s$, to be constant. In discrete
form, we solve:
\begin{eqnarray}
p_{0,j+1} &=& p_{0,j} + \frac{1}{2} \Delta r ( \rho_{0,j} + \rho_{0,j+1} ) g_{j+\myhalf},\\
s_{0,j+1} &=& s_{0,j},
\end{eqnarray}
with $\Delta r = 1.653125\times 10^{5}$~cm.  We begin with a guess of
$\rho_{0,j+1}$ and $T_{0,j+1}$ and use the equation of state and
Newton-Raphson iterations to find the values that satisfy our system.
Since this is a spherical, self-gravitating star, the gravitation
acceleration, $g_{j+\myhalf}$, is updated each iteration based on the
current value of the density. Once the temperature falls below
$10^7$~K, we keep the temperature constant, and continue determining
the density via hydrostatic equilibrium until the density falls to
$10^{-4}~\gcc$, after which we hold the density constant.  This
uniquely determines the initial model.  We note that this is the
same procedure we follow to initialize a convecting white dwarf 
for the multilevel low Mach number code, MAESTRO, described in 
\citet{multilevel}.

We map the model onto a ($5\times 10^8$~cm)$^3$ domain with 192$^3$,
384$^3$, and 768$^3$ grid cells, and center the star in the domain.
We let the simulation run to 1~s, and compare the maximum magnitude of
velocity vs.\ time and the magnitude of velocity vs.\ radius at $t=1$~s,  
a time greater than two sound-crossing times.
We only consider regions of the
star at $r<1.8\times 10^8$~cm, which corresponds to a density of
$\rho\approx 5.4\times 10^5~\gcc$.  Note that the density reaches the
floor of 10$^{-4}~\gcc$ at $r=1.9\times 10^8$~cm.  We turn on the
sponge at the radius where $\rho=100~\gcc$ and the sponge reaches its
full strength at the radius where $\rho=10^{-4}~\gcc$ with a sponge
strength of $\kappa=1000~\mathrm{s^{-1}}$.  We use a CFL of 0.9 and
no refinement.  We use the Helmholtz
equation of state \citep{timmes_swesty:2000,flash} and
no reactions are modeled.

\begin{figure}[t]
\centering
\includegraphics{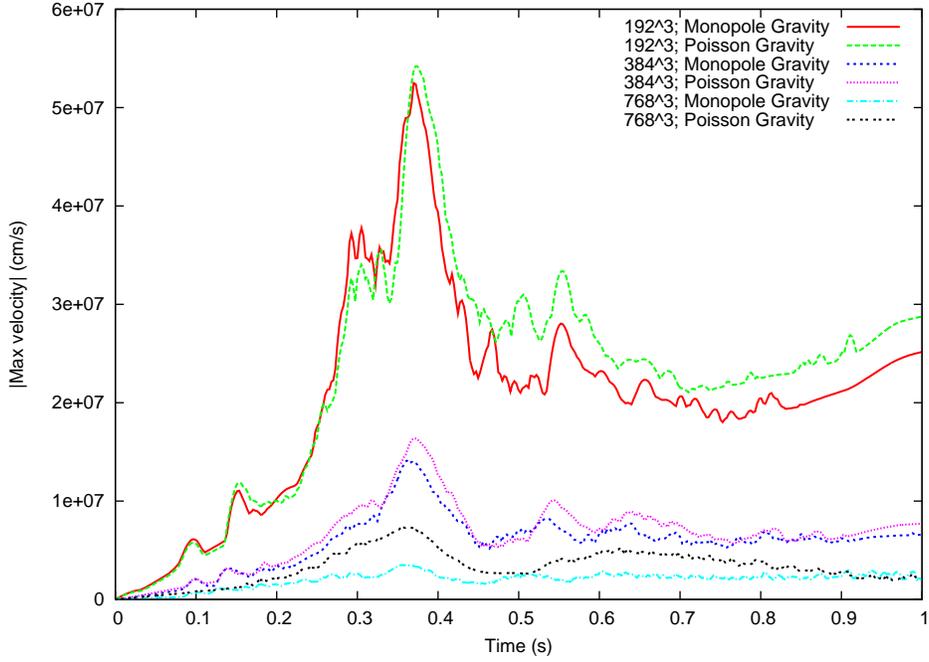}
\caption{\label{fig:StarGrav_UMAX} Maximum magnitude
of velocity vs.\ time for the stationary star gravity problem.  At each of
the three resolutions, we show the results using a monopole gravity
approximation and Poisson solve for gravity.  We note that in each
simulation, the maximum velocity is not strictly increasing, leading
us to believe that over longer periods of time the velocities will
remain small.  We note that sound speed at the center of the star is
approximately $9.4\times 10^8$~cm/s, so at the highest resolution, the
peak velocity is less than 1\% of the sound speed.  The solutions in
the monopole and Poisson cases match up very well; the discrepancy
we see at the finest resolution is due to large velocities at the edge
of the star, which is typically outside the region of interest.}
\end{figure}

Figure~\ref{fig:StarGrav_UMAX} shows a plot of the maximum magnitude
of velocity vs.\ time.  At each of
the three resolutions, we show the results using a monopole gravity
approximation and Poisson solve for gravity.  We note that in each
simulation, the maximum velocity is not strictly increasing, leading
us to believe that over longer periods of time the velocities will
remain small.  We note that sound speed at the center of the star is
approximately $9.4\times 10^8$~cm/s, so at the highest resolution, the
peak velocity is less than 1\% of the sound speed.  The monopole and
Poisson cases match up very well, except for the finest resolution.
The reason why we see larger peak velocities in the finest resolution
Poisson solver simulation is due to the large velocities at the edge
of the star.

\begin{figure}[t]
\centering
\includegraphics{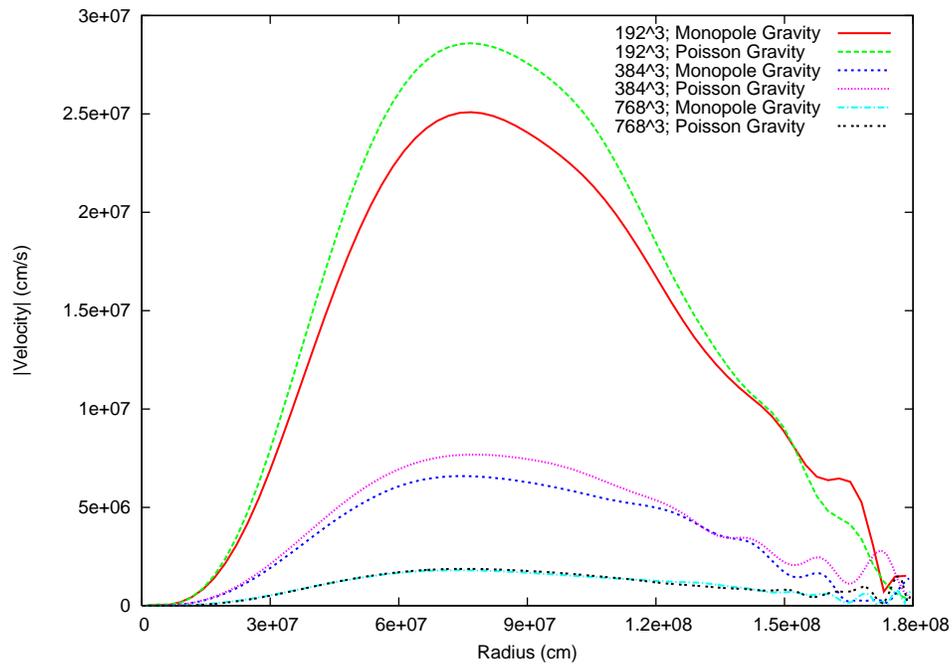}
\caption{\label{fig:StarGrav_profile} Magnitude of
velocity vs.\ radius at $t=1$~s for the stationary star gravity problem.
At each of the three resolutions, we show the results using a
monopole gravity approximation and Poisson solve for gravity.  Here,
we see clear second order convergence in the max norm, and the
monopole and Poisson simulations agree best at the highest resolution.}
%We also see how in the finest resolution runs, the velocities at the
%edge of the star can become large, but this is likely outside the
%region of interest for a typical simulation.
\end{figure}

Figure~\ref{fig:StarGrav_profile} shows a plot of the magnitude of
velocity vs.\ radius at $t=1$~s.
Again, at each of the three resolutions, we show the results using a
monopole gravity approximation and Poisson solve for gravity.  Here,
we see clear second order convergence in the max norm, and the
monopole and Poisson simulations agree best at the highest resolution.
We also see how in the finest resolution runs, the velocities at the
edge of the star can become large, but this is likely outside the
region of interest for a typical simulation.

\subsection{Homologous Dust Collapse}

As a second test of the gravity solver in CASTRO we implement the
homologous dust collapse test problem, a `pressure-less' configuration
that collapses under its own self-gravity.  An analytic solution that
describes the radius of the sphere as a function of time is found in
\citet{colgatewhite:1966}.  Our implementation of this problem follows
that described in \citet{flash_usersguide,monchmeyermuller}.  The problem is
initialized with a sphere with a large, uniform density, $\rho_0$, of radius
$r_0$.  The pressure everywhere should be negligible, i.e., the sound
crossing time should be much longer than the free-fall collapse time
(see, for example, \citealt{flash_usersguide}).
\citet{colgatewhite:1966} use $p = 0$.  We choose a value that does
not appear to affect the dynamics.  As the sphere collapses, the
density inside should remain spatially constant, but increase in value with
time.

Following \citet{flash_usersguide}, we take $\rho_0 = 10^9~\gcc$ and 
$r_0 = 6.5\times 10^8~\mathrm{cm}$.  The pressure is not specified, so
we take it to be $10^{15}$ dyn cm$^{-2}.$ Outside of the sphere, we 
set the density to $\rho_\mathrm{ambient} = 10^{-5}~\gcc$.  Finally,
since the sharp cutoff at the edge of the sphere is unphysical, we
smooth the initial profile by setting 
\begin{equation}
\rho = \rho_0 - \frac{\rho_0 - \rho_\mathrm{ambient}}{2}
   \left [ 1 + \tanh \left ( \frac{r - r_0}{h} \right ) \right ]
\end{equation}
with the smoothing length, $h = 4 \times 10^6 \ll r_0$. 
We use the gamma-law equation of state with $\gamma = 1.66$.

\begin{figure}[t]
\centering
\includegraphics{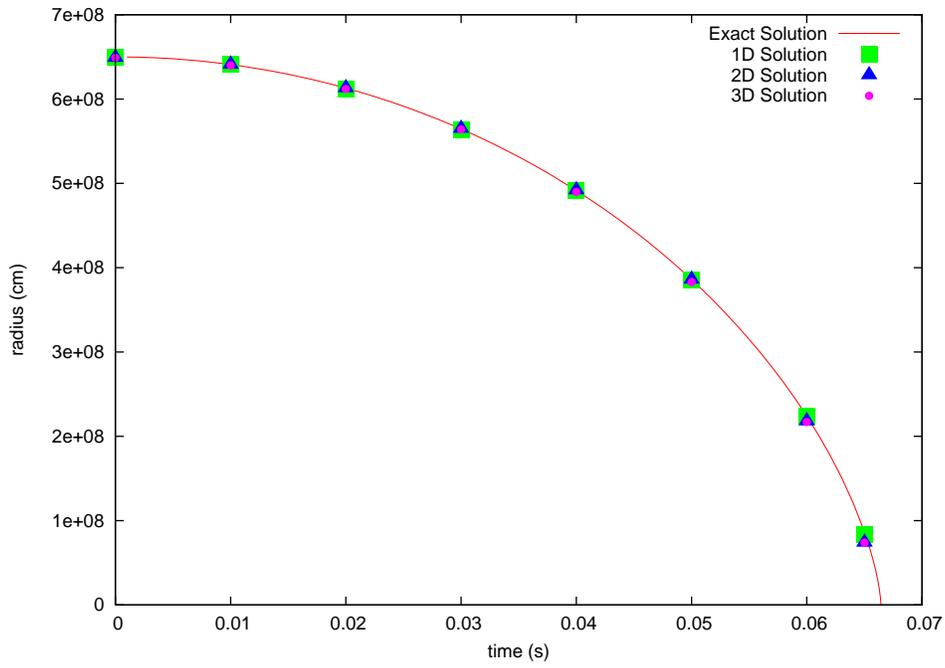}
\caption{\label{fig:dustcollapse} Radius vs.\ time for the homologous dust
collapse problem in 1D, 2D, and 3D simulations as compared to the exact solution.    
In all three cases we see excellent agreement with the exact solution. }
\end{figure}

Figure~\ref{fig:dustcollapse} shows the radius vs.\ time for the 
1D, 2D, and 3D simulations as compared to the exact solution.    
In all three cases we see excellent agreement with the exact solution.
%\MarginPar{Need to comment about density spike in the center of the star.}
%\MarginPar{For now not talking about that...}

\subsection{Reacting Bubbles in a Stellar Atmosphere}

A final test is a code comparison of the evolution of three reacting
bubbles in a plane-parallel stellar atmosphere.  This problem is
almost identical to the setup described in Section 4.2 of
\citet{ABNZ:III} with two minor differences.  First, we eliminate the
stably stratified layer at the base of the atmosphere by setting the
lower $y$ extrema of the domain to $5.00625\times 10^7$~cm---this way,
the bottommost row of cells in the domain is initialized with the
specified base density ($2.6\times 10^9~\gcc$) and temperature.
Second, we set the base temperature of the atmosphere to $6\times
10^8$~K (instead of $7\times 10^8$~K) to minimize the amount of
reactions occurring near the lower domain boundary.  Three
temperature perturbations are seeded in pressure-equilibrium with a
range of heights and widths as specified by equation (87) and Table 1 of
\cite{ABNZ:III}.  We use a uniform computation grid of $384 \times
576$ cells and a domain width of $2.16\times 10^8$~cm.

We compare the evolution to the FLASH code \citep{flash}, version 2.5,
using the standard dimensionally-split PPM hydrodynamics module that
comes with FLASH.  The lower boundary condition in both cases provides
hydrostatic support by integrating the equation of hydrostatic
equilibrium together with the equations of state into the ghost cells,
assuming a constant temperature, as described in \citet{ppm-hse}.  The
left and right boundary is periodic.  We
use the same single step ($^{12}\mathrm{C} + \, ^{12}\mathrm{C}
\rightarrow \, ^{24}\mathrm{Mg}$) reaction module described in
\citet{ABNZ:III}.  Both codes use the general stellar equation of
state described in \citet{flash,timmes_swesty:2000} with the Coulomb
corrections enabled.

\begin{figure}[t]
\centering
\includegraphics{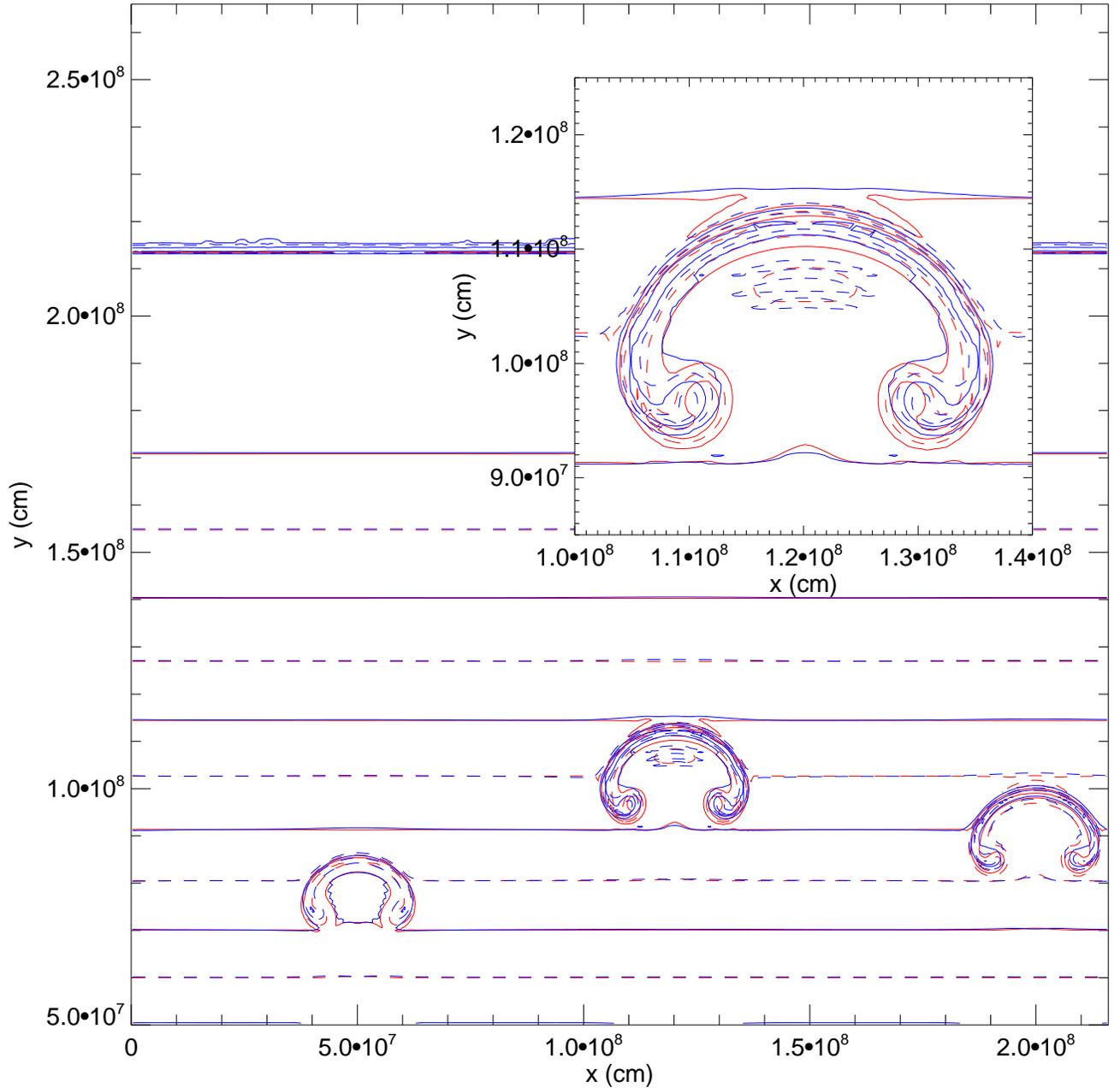}
\caption{\label{fig:test2_temp} Comparison of FLASH (red) and CASTRO
  (blue) temperature contours for the reacting bubble test.
  Temperature contours at $10^8$, $1.5\times 10^8$, $2\times 10^8$,
  $2.5\times 10^8$, $3.\times 10^8$, $3.5\times 10^8$, $4.\times 10^8$,
  $4.5\times 10^8$, $5.\times 10^8$, $5.5\times 10^8$, $6.\times 10^8$,
  $6.5\times 10^8$, $7.\times 10^8$, $7.5\times 10^8$, $8.\times 10^8$~K
  are shown, drawn with alternating solid and dashed lines.  The inset
  shows the detail of the middle bubble.  We see good
  agreement between FLASH and CASTRO.}
\end{figure}

\begin{figure}[t]
\centering
\includegraphics{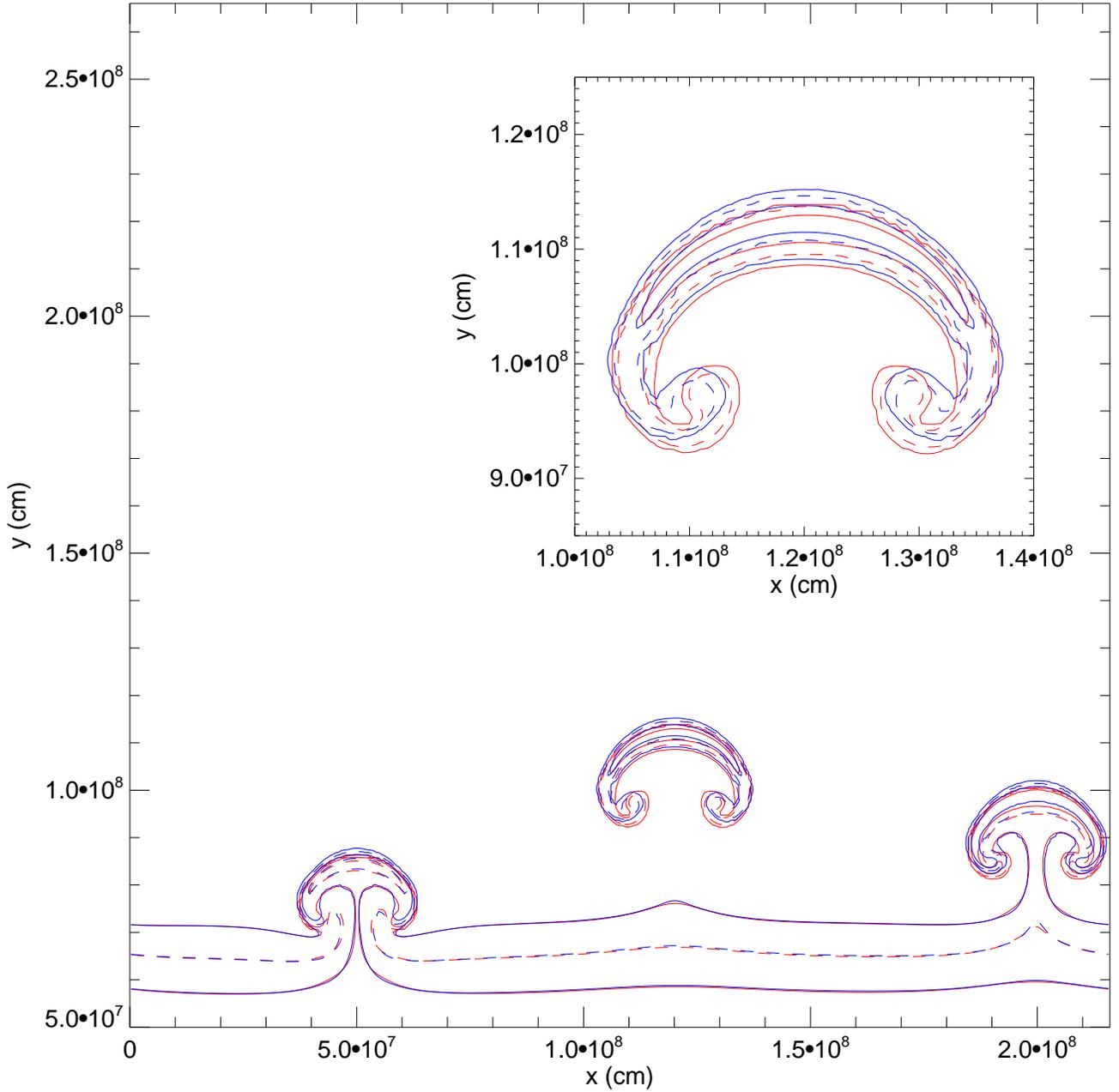}
\caption{\label{fig:test2_mg} Comparison of FLASH (red) and CASTRO
  (blue) $^{24}\mathrm{Mg}$ mass fraction contours for the reacting
  bubble test.  Contours are drawn at values of $X$ of $5\times
  10^{-9}$, $5\times 10^{-8}$, $5\times 10^{-7}$,$5\times 10^{-6}$,
  with alternating solid and dashed lines.  The inset shows the
  detail of the middle bubble.  As with the temperature,
  we see good agreement between FLASH and CASTRO.  }
\end{figure}

Figures~\ref{fig:test2_temp} and \ref{fig:test2_mg} show contours of
the temperature and $X(^{24}\mathrm{Mg})$ after 2.5~s of evolution for
both FLASH and CASTRO.  We see excellent agreement between the two
codes in terms of bubble heights and contour levels.

\subsection{Type Ia Supernova}

\begin{figure}[t]
\centering
\includegraphics{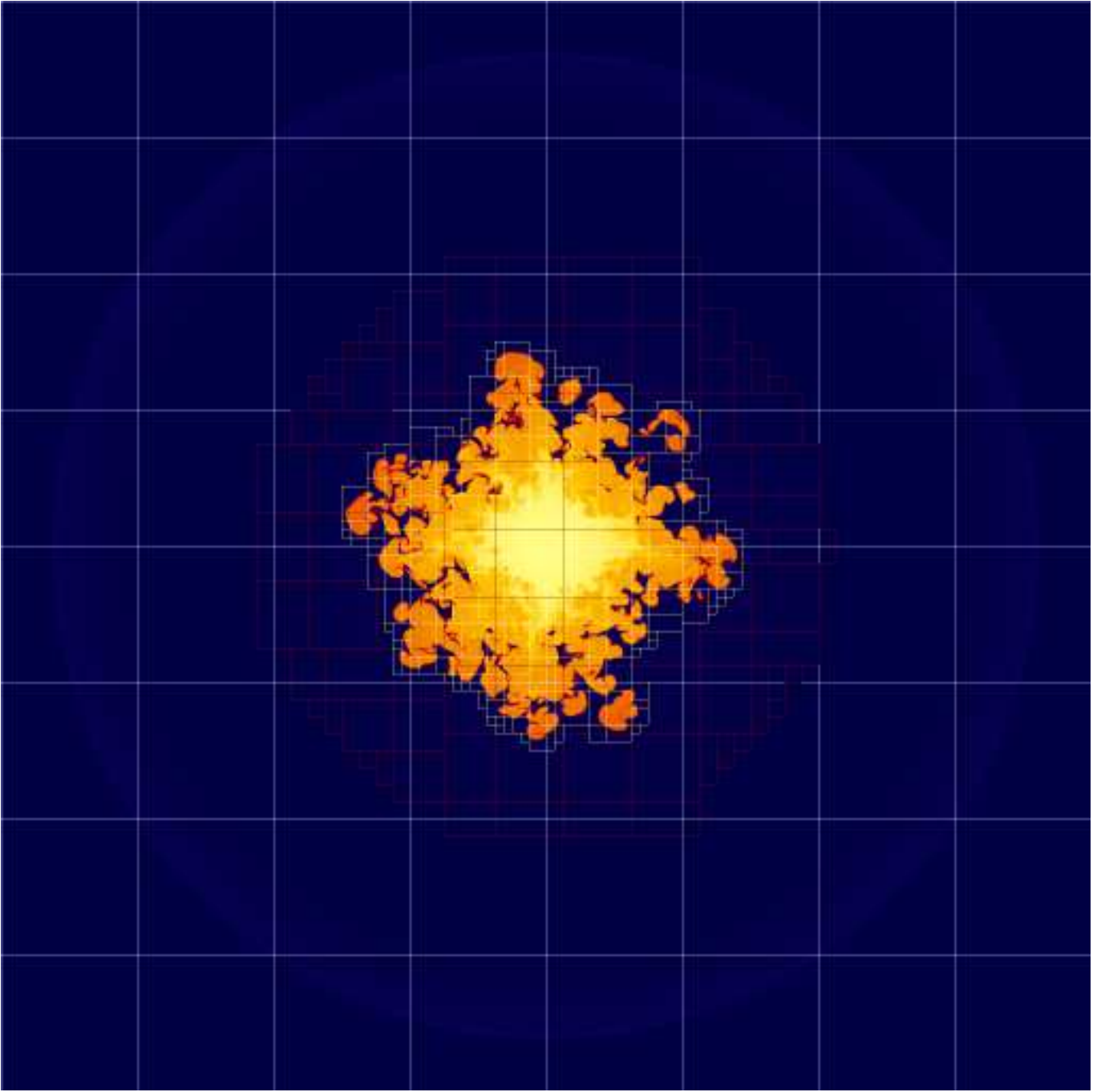}
\caption{\label{fig:3Dgrids_figure} Here we see a 2D slice of the temperature field from a 
3D calculation of a Type Ia supernova with two levels of refinement.  There are 512 grids,
each containing $64^3$ cells, at the coarsest level, over 1000 grids at level 1 and over 2000 grids at level 2.  Approximately
1.8\% of the domain is at the finest resolution.}
\end{figure}

\begin{figure}[t]
\centering
\includegraphics{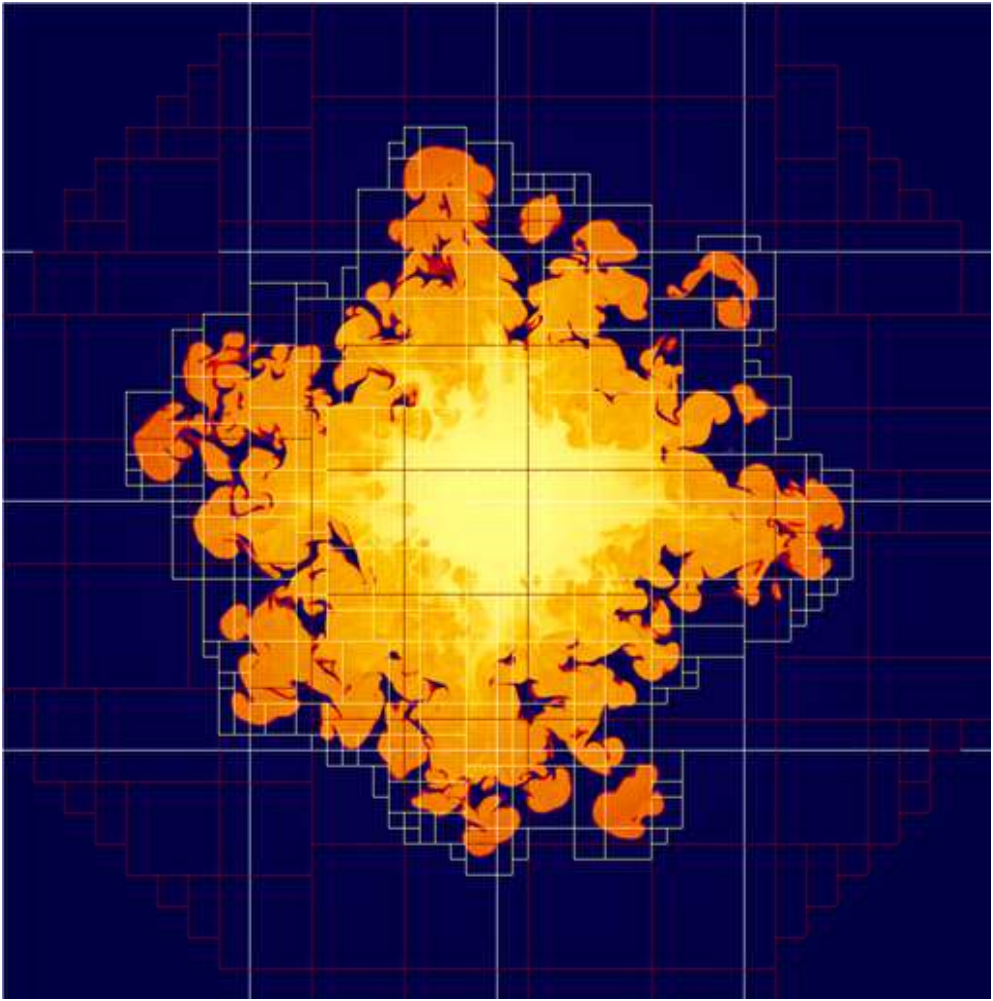}
\caption{\label{fig:3Dgrids_cropped} Here we see a close-up of the previous figure, showing more detail of the level 2 grids.}
\end{figure}

As a final example, in Figure~\ref{fig:3Dgrids_figure}
we show a 2D snapshot of temperature from a 3D calculation of a Type Ia 
supernova \citep{personal_ma_aja,ma_sneia}.  
This simulation uses a realistic stellar equation of state and a turbulent flame model, 
and is typical of more realistic CASTRO applications.   
The domain is 5.12 x $10^8$~cm on a side, 
and is covered with 512 $64^3$ grids.  There are two levels of factor two refinement, with 
approximately 1.8\% of the domain covered by level 2 grids with an effective resolution of
2.5 x $10^5$~cm.  Figure~\ref{fig:3Dgrids_cropped} is a close-up of the center of the
domain so that the level 2 grids are more visible.

\section{Summary}

We have described a new Eulerian adaptive mesh code, CASTRO, for
solving the multicomponent compressible hydrodynamic equations with a
general equation of state for astrophysical flows.  CASTRO differs
from existing codes of its type in that it uses unsplit PPM for its
hydrodynamic evolution, subcycling in time, and a nested hierarchy of
logically-rectangular grids.  Additional physics
includes self-gravitation, nuclear reactions, and radiation.
Radiation will be described in detail in the next paper, Part II, of
this series.

CASTRO is currently being used in simulations of Type Ia supernovae
and core-collapse supernovae; examples of simulations done using
CASTRO can be found in \citet{Joggerstetal:2009,woosley-scidac2009}.
Further details on the CASTRO algorithm can be found in the CASTRO
User Guide \citep{CASTROUserGuide}.

\acknowledgements

We thank Alan Calder for useful discussions on test problems
and Stan Woosley for numerous invaluable interactions.
In comparing to other codes, we benefited from helpful discussions
with Brian O'Shea about Enzo, Paul Ricker about gravity in FLASH,
and Michael Clover about RAGE.
Finally, we thank Haitao Ma, Jason Nordhaus and Ken Chen for being 
patient early users of CASTRO.
The work at LBNL was supported by the Office of High Energy Physics 
and the Office of Mathematics, Information, and Computational Sciences
as part of
the SciDAC Program under the U.S. Department of Energy under contract
No.\ DE-AC02-05CH11231.  
The work performed at LLNL was under the auspices of the U.S.
Department of Energy under contract No.\ DE-AC52-07NA27344.
MZ was supported by Lawrence Livermore
National Lab under contracts B568673, B574691, and B582735.
This research used resources of the National Energy Research Scientific 
Computing Center, which is supported by the Office of Science of the U.S. 
Department of Energy under Contract No. DE-AC02-05CH11231.
This research used resources of the Oak Ridge Leadership Computational
Facility (OLCF), which is supported by the 
Office of Science of the Department of Energy under Contract DE-AC05-00OR22725.

\clearpage

\bibliographystyle{apj}
\bibliography{ws}

\begin{thebibliography}{51}
\expandafter\ifx\csname natexlab\endcsname\relax\def\natexlab#1{#1}\fi

\bibitem[{Almgren {et~al.}(1998)Almgren, Bell, Colella, Howell, \&
  Welcome}]{almgren-iamr}
Almgren, A.~S., Bell, J.~B., Colella, P., Howell, L.~H., \& Welcome, M.~L.
  1998, Journal of Computational Physics, 142, 1

\bibitem[{{Almgren} {et~al.}(2008){Almgren}, {Bell}, {Nonaka}, \&
  {Zingale}}]{ABNZ:III}
{Almgren}, A.~S., {Bell}, J.~B., {Nonaka}, A., \& {Zingale}, M. 2008,
  Astrophysical Journal, 684, 449

\bibitem[{Bell {et~al.}(1994)Bell, Berger, Saltzman, \& Welcome}]{bell-3d}
Bell, J., Berger, M., Saltzman, J., \& Welcome, M. 1994, SIAM J. Sci. Statist.
  Comput., 15, 127

\bibitem[{{Bell} {et~al.}(1989){Bell}, {Colella}, \&
  {Trangenstein}}]{bellcolellatrangenstein}
{Bell}, J.~B., {Colella}, P., \& {Trangenstein}, J.~A. 1989, Journal of
  Computational Physics, 82, 362

\bibitem[{Berger \& Colella(1989)}]{berger-colella}
Berger, M.~J., \& Colella, P. 1989, Journal of Computational Physics, 82, 64

\bibitem[{Berger \& Oliger(1984)}]{berger-oliger}
Berger, M.~J., \& Oliger, J. 1984, Journal of Computational Physics, 53, 484

\bibitem[{Berger \& Rigoutsos(1991)}]{bergerRigoutsos:1991}
Berger, M.~J., \& Rigoutsos, J. 1991, IEEESMC, 21, 1278

\bibitem[{{Bryan} {et~al.}(1995){Bryan}, {Norman}, {Stone}, {Cen}, \&
  {Ostriker}}]{bryan:1995}
{Bryan}, G.~L., {Norman}, M.~L., {Stone}, J.~M., {Cen}, R., \& {Ostriker},
  J.~P. 1995, Computer Physics Communications, 89, 149

\bibitem[{{CASTRO User Guide}(2009)}]{CASTROUserGuide}
{CASTRO User Guide}. 2009,
  \texttt{https:ccse.lbl.gov/Research/CASTRO/CastroUserGuide.pdf}

\bibitem[{{Chandrasekhar}(1961)}]{chandrasekharbook}
{Chandrasekhar}, S. 1961, {Hydrodynamic and hydromagnetic stability}, ed.
  S.~Chandrasekhar, dover reprint, 1981

\bibitem[{{Colella}(1990)}]{colella1990}
{Colella}, P. 1990, Journal of Computational Physics, 87, 171

\bibitem[{{Colella} \& {Glaz}(1985)}]{colellaglaz1985}
{Colella}, P., \& {Glaz}, H.~M. 1985, Journal of Computational Physics, 59, 264

\bibitem[{{Colella} {et~al.}(1997){Colella}, {Glaz}, \& {Ferguson}}]{cgf}
{Colella}, P., {Glaz}, H.~M., \& {Ferguson}, R.~E. 1997, unpublished manuscript

\bibitem[{Colella \& Sekora(2008)}]{ppm2}
Colella, P., \& Sekora, M.~D. 2008, Journal of Computational Physics, 227, 7069

\bibitem[{Colella \& Woodward(1984)}]{ppm}
Colella, P., \& Woodward, P.~R. 1984, Journal of Computational Physics, 54, 174

\bibitem[{{Colgate} \& {White}(1966)}]{colgatewhite:1966}
{Colgate}, S.~A., \& {White}, R.~H. 1966, \apj, 143, 626

\bibitem[{Crutchfield(1991)}]{crutchfield:1991}
Crutchfield, W.~Y. 1991, Load Balancing Irregular Algorithms, Tech. Rep.
  UCRL-JC-107679, LLNL

\bibitem[{{Dimonte} {et~al.}(2004){Dimonte}, {Youngs}, {Dimits}, {Weber},
  {Marinak}, {Wunsch}, {Garasi}, {Robinson}, {Andrews}, {Ramaprabhu}, {Calder},
  {Fryxell}, {Biello}, {Dursi}, {MacNeice}, {Olson}, {Ricker}, {Rosner},
  {Timmes}, {Tufo}, {Young}, \& {Zingale}}]{RTalpha}
{Dimonte}, G. {et~al.} 2004, Physics of Fluids, 16, 1668

\bibitem[{{FLASH 3.2 User's Guide}(2009)}]{flash_usersguide}
{FLASH 3.2 User's Guide}. 2009, http://flash.uchicago.edu/website/codesupport/

\bibitem[{{Franklin Performance Monitoring}(2010)}]{nersc_io}
{Franklin Performance Monitoring}. 2010, {N5 IOR Aggregate Write},
  \url{http://www.nersc.gov/nusers/systems/franklin/monitor.php}

\bibitem[{{Fryxell} {et~al.}(2000){Fryxell}, {Olson}, {Ricker}, {Timmes},
  {Zingale}, {Lamb}, {MacNeice}, {Rosner}, {Truran}, \& {Tufo}}]{flash}
{Fryxell}, B. {et~al.} 2000, Astrophysical Journal Supplement, 131, 273

\bibitem[{{Gittings} {et~al.}(2008){Gittings}, {Weaver}, {Clover}, {Betlach},
  {Byrne}, {Coker}, {Dendy}, {Hueckstaedt}, {New}, {Oakes}, {Ranta}, \&
  {Stefan}}]{RAGE}
{Gittings}, M. {et~al.} 2008, Computational Science and Discovery, 1, 015005

\bibitem[{Joggerst {et~al.}(2009)Joggerst, Almgren, Bell, Heger, Whalen, \&
  Woosley}]{Joggerstetal:2009}
Joggerst, C.~C., Almgren, A., Bell, J., Heger, A., Whalen, D., \& Woosley,
  S.~E. 2009, Astrophysical Journal

\bibitem[{{Kamm} \& {Timmes}(2007)}]{timmes_sedov_code}
{Kamm}, J.~R., \& {Timmes}, F.~X. 2007, submitted to ApJ supplement, May 2007,
  see \texttt{http://cococubed.asu.edu/code\_pages/sedov.shtml}

\bibitem[{{Lattimer} \& {Swesty}(1991)}]{LSEOS}
{Lattimer}, J.~M., \& {Swesty}, F.~D. 1991, Nuclear Physics A, 535, 331, code
  obtained from \texttt{http://www.astro.sunysb.edu/dswesty/lseos.html}

\bibitem[{{Layzer}(1955)}]{layzer:1955}
{Layzer}, D. 1955, \apj, 122, 1

\bibitem[{Ma \& Aspden(2010)}]{personal_ma_aja}
Ma, H., \& Aspden, A.~J. 2010, Private communication

\bibitem[{{Ma} {et~al.}(2010){Ma}, {Woosley}, {Almgren}, \& {Bell}}]{ma_sneia}
{Ma}, H., {Woosley}, S., {Almgren}, A., \& {Bell}, J. 2010, in American
  Astronomical Society Meeting Abstracts, Vol. 215, American Astronomical
  Society Meeting Abstracts, 343.01--+

\bibitem[{McCorquodale \& Colella(2010)}]{ppm3}
McCorquodale, P., \& Colella, P. 2010, Journal of Computational Physics, to
  appear

\bibitem[{{Miller} \& {Colella}(2002)}]{ppmunsplit}
{Miller}, G.~H., \& {Colella}, P. 2002, Journal of Computational Physics, 183,
  26

\bibitem[{Miniati \& Colella(2007)}]{miniati-colella}
Miniati, F., \& Colella, P. 2007, Journal of Computational Physics, 227, 400

\bibitem[{{Monchmeyer} \& {Muller}(1989)}]{monchmeyermuller}
{Monchmeyer}, R., \& {Muller}, E. 1989, \aap, 217, 351

\bibitem[{{M\"uller}(1986)}]{mueller:1986}
{M\"uller}, E. 1986, \aap, 162, 103

\bibitem[{{Nonaka} {et~al.}(2010){Nonaka}, {Almgren}, {Bell}, {Lijewski},
  {Malone}, \& {Zingale}}]{multilevel}
{Nonaka}, A., {Almgren}, A.~S., {Bell}, J.~B., {Lijewski}, M.~J., {Malone}, C.,
  \& {Zingale}, M. 2010, ApJS, submitted

\bibitem[{{Omang} {et~al.}(2006){Omang}, {B{\o}rve}, \& {Trulsen}}]{omang:2006}
{Omang}, M., {B{\o}rve}, S., \& {Trulsen}, J. 2006, Journal of Computational
  Physics, 213, 391

\bibitem[{{O'Shea} {et~al.}(2005){O'Shea}, {Bryan}, {Bordner}, {Norman},
  {Abel}, {Harkness}, \& {Kritsuk}}]{ENZO}
{O'Shea}, B.~W., {Bryan}, G., {Bordner}, J., {Norman}, M.~L., {Abel}, T.,
  {Harkness}, R., \& {Kritsuk}, A. 2005, in {Lecture Notes in Computational
  Science and Engineering}, Vol.~41, {Adaptive Mesh Refinement -- Theory and
  Applications}, ed. T.~{Plewa}, T.~{Linde}, \& V.~G. {Weirs} (Springer),
  341--350

\bibitem[{Plewa \& M\"uller(1999)}]{PlewaMueller:1999}
Plewa, T., \& M\"uller, E. 1999, Astronomy and Astrophysics, 342, 179

\bibitem[{Rendleman {et~al.}(2000)Rendleman, Beckner, Lijewski, Crutchfield, \&
  Bell}]{rendleman-hyper}
Rendleman, C.~A., Beckner, V.~E., Lijewski, M., Crutchfield, W.~Y., \& Bell,
  J.~B. 2000, Computing and Visualization in Science, 3, 147

\bibitem[{{Ricker}(2008)}]{ricker_multigrid:2008}
{Ricker}, P.~M. 2008, \apjs, 176, 293

\bibitem[{Saltzman(1994)}]{saltzman}
Saltzman, J. 1994, Journal of Computational Physics, 115, 153

\bibitem[{{Sedov}(1959)}]{sedov:1959}
{Sedov}, L.~I. 1959, Similarity and Dimensional Methods in Mechanics (Academic
  Press), translated from the 4th Russian Ed.

\bibitem[{{Sharp}(1984)}]{sharp:1984}
{Sharp}, D.~H. 1984, Physica D Nonlinear Phenomena, 12, 3

\bibitem[{{Sod}(1978)}]{sod:1978}
{Sod}, G.~A. 1978, Journal of Computational Physics, 27, 1

\bibitem[{{Strang}(1968)}]{strang:1968}
{Strang}, G. 1968, {SIAM J. Numerical Analysis}, 5, 506

\bibitem[{{Taylor}(1950)}]{taylor:1950}
{Taylor}, G. 1950, Royal Society of London Proceedings Series A, 201, 192

\bibitem[{Timmes \& Swesty(2000)}]{timmes_swesty:2000}
Timmes, F.~X., \& Swesty, F.~D. 2000, Astrophysical Journal Supplement, 126,
  501

\bibitem[{Toro(1997)}]{Toro}
Toro, E.~F. 1997, Riemann Solvers and Numerical Methods for Fluid Dynamics
  (Springer)

\bibitem[{{VisIt User's Manual}(2005)}]{visit}
{VisIt User's Manual}. 2005,
  \texttt{https://wci.llnl.gov/codes/visit/home.html}

\bibitem[{{Woosley} {et~al.}(2009){Woosley}, {Almgren}, {Aspden}, {Bell},
  {Kasen}, {Kerstein}, {Ma}, {Nonaka}, \& {Zingale}}]{woosley-scidac2009}
{Woosley}, S.~E. {et~al.} 2009, Journal of Physics Conference Series, 180,
  012023

\bibitem[{Zingale {et~al.}(2009)Zingale, Almgren, Bell, Nonaka, \&
  Woosley}]{ZABNW:IV}
Zingale, M., Almgren, A.~S., Bell, J.~B., Nonaka, A., \& Woosley, S.~E. 2009,
  Astrophysical Journal, 704, 196

\bibitem[{{Zingale} {et~al.}(2002){Zingale}, {Dursi}, {ZuHone}, {Calder},
  {Fryxell}, {Plewa}, {Truran}, {Caceres}, {Olson}, {Ricker}, {Riley},
  {Rosner}, {Siegel}, {Timmes}, \& {Vladimirova}}]{ppm-hse}
{Zingale}, M. {et~al.} 2002, Astrophysical Journal Supplement, 143, 539

\end{thebibliography}

\clearpage

\clearpage

\clearpage

\clearpage

\clearpage

\clearpage

\clearpage

\clearpage

\clearpage

\clearpage

\clearpage

\clearpage

\clearpage

\clearpage

\clearpage

\end{document}